\documentclass[12pt, preprint,aps,prc,showpacs,tightenlines]{revtex4}
\usepackage{epsf}

\def\beq{\begin{equation}}
\def\eeq{\end{equation}}
\def\bea{\begin{eqnarray}}
\def\eea{\end{eqnarray}} 
\def\beqa{\begin{equation}\begin{array}{l}}
\def\eeqa{\end{array}\end{equation}}
\def\eqlab#1{\label{eq:#1}}

\def\eref#1{(\ref{eq:#1})}
\def\Eqref#1{Eq.~(\ref{eq:#1})}

\def\slap{p \hspace{-2mm} \slash}

\def\half{\mbox{\small{$\frac{1}{2}$}}}

\def\quarter{\mbox{\small{$\frac{1}{4}$}}}
\def\third{\mbox{\small{$\frac{1}{3}$}}}

\def\barr{\left(\begin{array}{c}}
\def\earr{\end{array}\right)}
\def\bmat{\left(\begin{array}{cc}}
\def\emat{\end{array}\right)}
\def\al{\alpha}
\def\be{\beta}
\def\ga{\gamma} \def\Ga{{\it\Gamma}}
\def\de{\delta} \def\De{\Delta}
\def\veps{\varepsilon}  \def\eps{\epsilon}

\def\la{\lambda} \def\La{{\Lambda}}

 \def\Si{{\it\Sigma}}
\def\th{\theta}  
\def\w{\omega}

\def\pa{\partial}

\def\barf{\zeta}
\def\Z{{\mathcal Z}}

\def\pa{\partial}

\def\nn{\nonumber}

\def\lag{{\mathcal L}}

\def\mathscr{\mathcal}

\def\3d{3-D}

\def\ODI{O$\Delta$I}
\def\ODR{O$\Delta$R}
\newcommand{\lsim}{\, \, \raisebox{-0.8ex}{$\stackrel{\textstyle <}{\sim}$ }}
\def\ol#1{\overline{#1}} 

\begin{document}

\title{Effective theory of the $\Delta$(1232) in
 Compton scattering \\
off the nucleon}

\author{Vladimir Pascalutsa}
\email{vlad@phy.ohiou.edu}
\author{Daniel R. Phillips}
\email{phillips@phy.ohiou.edu}

\affiliation{
Department of Physics and Astronomy, Ohio University,
Athens, OH 45701}
\date{\today}

\begin{abstract}
We formulate a new power-counting scheme for a chiral effective field
theory of nucleons, pions, and Deltas. This extends chiral
perturbation theory into the Delta-resonance region.  We calculate
nucleon Compton scattering up to next-to-leading order in this
theory. The resultant description of existing $\gamma$p cross
section data is very good for photon energies up to about 300 MeV.  We
also find reasonable numbers for the spin-independent polarizabilities
$\alpha_p$ and $\beta_p$.
\end{abstract}

\maketitle
\thispagestyle{empty}

\section{Introduction}

Compton scattering on the proton ($\gamma$p) and the deuteron ($\ga$D)
provides a clean and unique probe of nucleon electromagnetic
structure, revealing information different to that obtained in
electron scattering.  During the past decade a number of excellent
experimental programs have been dedicated to these two processes (see
Refs.~\cite{Hal93,MacG95,LEGS97,LEGS01,MAMI01} and
\cite{Lu94,Ho99,Lu02,Ko00,Ko02}, respectively). At low photon
energies, these experiments probe the static properties of the
nucleon, such as its electric charge, magnetic moment, and
polarizabilities. Above the pion-production threshold, the process
becomes dominated by the excitation of resonances, most prominently
the $\De$(1232)--isobar.  Many theoretical methods aim at
understanding this process in both the low-energy and the resonance
region. In particular, significant progress has been made recently
using dispersion relations~\cite{DR,DR2} and effective Lagrangian
models~\cite{PaS95,Giessen,KoS01}.  On the other hand, previous
calculations using chiral perturbation theory ($\chi$PT) appear to
work only at low photon energies---energies at or below the
pion-production threshold~\cite{Bab97,McG01}.  This present study
attempts to extend these $\chi$PT calculations above the pion
threshold and into the Delta-resonance region.

In the low-energy regime $\chi$PT seems to work extremely well.
 At next-to-leading order (NLO), i.e.,
third order in small momenta [$=O(q^3)$], heavy-baryon (HB) $\chi$PT
predicts, for the electric and magnetic polarizabilities~\cite{Be93,BKM}:
\begin{eqnarray}
\eqlab{alphabet}
&&\alpha_p=\alpha_n=\frac{5 \pi \al }{6\, m_\pi} \left( \frac{g_A}{4
\pi f_\pi}\right)^2
=12.2 \times 10^{-4}~\mbox{fm}^3,\\ 
&&\beta_p=\beta_n=\frac{\pi \al}{12\, m_\pi} 
\left( \frac{g_A}{4 \pi f_\pi}\right)^2 =1.2 \times 10^{-4}~\mbox{fm}^3\,,
\end{eqnarray}
where $\al=e^2/4\pi\simeq 1/137$, $g_A\simeq 1.26$, $f_\pi \simeq 93$
MeV, $m_\pi \simeq 139$ MeV~\footnote{Throughout this paper the
designations LO, NLO, etc. refer to the order in the $\gamma$N
amplitude. These one-loop results are, strictly speaking, {\it
leading}-order predictions for $\alpha_p$ and $\beta_p$, but we
refer to them as next-to-leading order (NLO) since Eq.~(\ref{eq:alphabet}) 
is derived by considering the NLO result for the nucleon Compton amplitude.}.
Since there are no Compton counter-terms present at
$O(q^3)$, this is a genuine prediction of $\chi$PT. A prediction
which---at least for the proton---is in remarkable agreement with
recent extractions of these quantities from low-energy data,
e.g.~\cite{Be02}: 
\begin{eqnarray}
\alpha_p&=& (12.1\pm 1.1 \pm 0.5) \times 10^{-4}~{\rm fm}^3,\\
\beta_p&=& (3.2\pm 1.1 \pm 0.1) \times 10^{-4}~{\rm fm}^3,
\end{eqnarray}
where the first error is statistical and the second represents
the theory error of the fit to data.

However, the agreement of the NLO HB$\chi$PT prediction with the
experimental $\ga$p cross-section data is good only up to photon
energies $\w\simeq 100$ MeV~\cite{Bab97}. The recent NNLO [$O(q^4)$]
calculation~\cite{McG01} agrees with experiment to slightly higher
energies, but above $\w \simeq 120$ MeV significant discrepancies
begin to appear, most notably at backward angles.  
This is
perhaps not surprising, since the Delta-isobar excitation is not
included explicitly in this chiral expansion. And, as we shall argue,
the breakdown scale of $\chi$PT without an explicit Delta is set
essentially by the Delta-nucleon mass difference:
\beq
\De \equiv
M_\De - M_N\approx 293\, 
\mbox{MeV} \,.
\eeq
Thus, to extend the region of $\chi$PT applicability to 
$\w \sim \De $, the Delta must be included as an explicit degree
of freedom.

The Delta contribution for the Compton amplitude had already been
analyzed using chiral effective Lagrangians with explicit Deltas in
Refs.~\cite{But92,Ber93,He97}.  These studies focused mainly on
nucleon polarizabilities. The predictions made in
Refs.~\cite{But92,Ber93,BKM} are obscured by off-shell ambiguities, in
particular by the so-called {\it ``off-shell parameters''} which
control the infamous spin-1/2 sector of the spin-3/2 Delta field. In a
``reasonable'' range for these parameters the Delta contribution to
$\beta_p^{(\De)}$ varies between 0 and $14
\times 10^{-4}$~fm$^3$~\cite{Ber93}.  In contrast, Hemmert {\it et.~al.} 
\cite{He97}, to next-to-leading order in their {\it small scale expansion}
(SSE)~\cite{He98}, find a result which is independent of the off-shell
parameters, and thus is apparently a reliable prediction. But this
prediction for the Delta-contribution to the magnetic polarizability
is $\be^{(\De)}_p\approx 9\times 10^{-4}$~fm$^3$, in dramatic
contradiction with experiment~\cite{He98B}.  (For an attempt to
remedy this using a ``modified SSE'' see Ref.~\cite{Gr02}.)

In this work we include the Delta in the chiral Lagrangian in a
fashion somewhat different to this literature. First of all, the
Lagrangian is written such that the unphysical spin-1/2 components of
the Delta field decouple from observables~\cite{Pa98,PaT99}, hence no
``off-shell parameters'' appear. This feature of our Lagrangian,
besides removing the redundant parameters, allows us to dress the
Delta-pole contribution in a manifestly covariant way.

Furthermore, we set up our power-counting scheme so that it is both closely
connected to the usual $\chi$PT without explicit Deltas in the
low-energy region $\w\sim m_\pi$, {\it and} extends to the Delta region
$\w\sim \De$. This is achieved by recognizing the hierarchy of scales:
\beq
\eqlab{scalehi}
m_\pi \ll \De\ll \La \sim 1 \,\,\,\mbox{GeV}\,,
\eeq
where $\La$ stands for the ``high-energy scale'', the breakdown scale
of the theory. Therefore, our scheme is rather different from the SSE
of Refs.~\cite{He97,He98} (see also Ref.~\cite{JM91}), where the
Delta-nucleon mass-difference is assumed to be of order $m_\pi$ (i.e.,
$\De\sim m_\pi$).  A more detailed comparison of our scheme and the
SSE will be given below.

With the three-scale hierarchy~\eref{scalehi} one in principle has
two small expansion parameters: 
$m_\pi/\De$ and $\De/\La$.
We regard both of them as roughly the same size and so introduce a
single small parameter:
\begin{equation}
\delta = \frac{\Delta}{\Lambda} \sim \frac{m_\pi}{\Delta}.
\end{equation}
Note that this implies that $m_\pi$ scales as $\de^2$.

The validity of the scale hierarchy~\eref{scalehi} and the expansion
in powers of $\de$ (which we shall refer to as the {\it
$\de$-expansion}) is to be judged by the success of the resultant
effective-field-theory (EFT) description of processes involving the
excitation of Delta. We regard the results we shall present here for
$\ga$p scattering as significant evidence in favor of this EFT
expansion.

To obtain the NLO result for $\gamma$N scattering in our scheme for
both the low-energy and the Delta regions, the Delta-pole contribution
to this process must be dressed, and then added to the NLO HB$\chi$PT
result.  This introduces two free parameters which characterize the
strength of the $\ga N \rightarrow \De$ transition: $g_M$ and $g_E$.
Adjusting these parameters we find very good agreement with the
experimental $\ga \mbox{p}$ differential cross section up to $\w
\approx 300$ MeV, thereby extending the domain
of applicability of chiral EFT into the Delta region. At the same
time we also find reasonable values for the nucleon polarizabilities.

In the next section we introduce the Lagrangian for the Delta
and discuss its properties.  Section~\ref{sec-pc} then describes the
$\delta$-expansion for Compton scattering on the proton. In
particular, we show that for $\omega \sim m_\pi$ the power counting is
very similar to that of HB$\chi$PT, while for $\omega \sim \Delta$ the
power counting mandates resummation of the $\Delta$ propagator,
thereby dressing the $\Delta$ and giving it a finite width. In
Section~\ref{sec-practice} we summarize the elements of our
calculation, and then in Section~\ref{sec-results} we present and
discuss the results of our NLO calculation for the differential
cross section, as well as for the spin-independent polarizabilities
$\alpha_p$, $\beta_p$.  We conclude in Section~\ref{sec-conc}.

\section{The Chiral Lagrangian}

The pion-nucleon sector of the HB$\chi$PT Lagrangian is well discussed
in the literature, see e.g.~\cite{BKM}. The terms relevant for our
purposes are%
~\footnote{Our conventions: metric tensor
$g^{\mu\nu}={\rm diag}(1,-1,-1,-1)$; $\ga$-matrices
$\ga^\mu$, $\ga^5=i\ga^0\ga^1\ga^2\ga^3$,
$\{\ga^\mu,\ga^\nu\}
=2g^{\mu\nu}$;
fully antisymmetrized products of $\ga$-matrices 
$\ga^{\mu\nu}=\half[\ga^\mu,\ga^\nu]=\ga^\mu\ga^\nu-g^{\mu\nu}$,
$\ga^{\mu\nu\al}=\half \{\ga^{\mu\nu},\ga^\al\}=
i\veps^{\mu\nu\al\be}\ga_\be\ga_5$,
$\ga^{\mu\nu\al\be}=\half [\ga^{\mu\nu\al},\ga^\be]=
i\veps^{\mu\nu\al\be}\ga_5$; spinor indices are omitted.}
:
\bea
\eqlab{piNlag}
\lag &=& \half D_\mu^{ab} \pi_a\, D^{\mu\, bc} \pi_c  - \half 
m_\pi^2 \pi^2 
+\ol N\, \left[ i \ga\cdot D -M_N - \frac{g_A}{2f_\pi}  
(\ga\cdot D\, \pi_a) \tau_a \gamma_5
+ \frac{\kappa}{4M_N} \ga^{\mu\nu} F_{\mu\nu} \right ]\,N \, \nonumber\\
&& -\frac{e^2 }{32\pi^2 f_\pi} F_{\mu\nu} \tilde F^{\mu\nu} \pi_3+ \ldots,
\label{eq:pinlag}
\eea
where $\pi_a$ represents the isovector pseudoscalar pion
field, $N$ is the isodoublet spinor field of the nucleon, $\tau_a$ are
the isospin Pauli matrices, $D_\mu = \pa_\mu - ieQ A_\mu$, with $Q$
representing the electric charge isospin operator ($Q_\pi = -i\veps^{ab3}$,
$Q_N=\half(1+\tau_3)$),
$A_\mu$ the electromagnetic field,
$F_{\mu\nu}=\pa_\mu A_\nu -\pa_\nu A_\mu $, $\tilde F^{\mu\nu}= \half
\veps^{\mu\nu\al\be} F_{\al\be}$, and $\kappa$ is the
anomalous magnetic moment of the nucleon ($\kappa_p \simeq 1.79$,
$\kappa_n\simeq -1.91$).

Next we specify the terms involving
the Delta field. 
Describing the Delta field by an isospin-3/2 spin-3/2 Rarita-Schwinger 
(RS) vector-spinor $\De_\mu(x)$, 
we write the Delta piece of the chiral Lagrangian in the following form:
\bea
\eqlab{Delag}
\lag &=& \lag_{RS} +  \lag_{\pi N\De} +\lag_{\ga N \De} +\ldots \,,
\\
\eqlab{freeRS}
\lag_{RS} &=& \ol\De_\mu \left(i\ga^{\mu\nu\al}\,\pa_\al - 
M_\De\,\ga^{\mu\nu}\right) \De_\nu\, ,
\\
\eqlab{pinDe}
\lag_{\pi N\De}&=& \frac{i h_A}{2 f_\pi M_\De}\,
\ol N\, T_a^\dagger \,\ga^{\mu\nu\la}\, (\pa_\mu \De_\nu)\, \pa_\la \pi^a 
+ \mbox{H.c.}\, ,\\
\eqlab{ganDe}
\lag_{\ga N \De}&=& \frac{3\,e}{2M_N(M_N+M_\Delta)}\,\ol N\, T_3^\dagger
\left(i g_M  \tilde F^{\mu\nu}
- g_E \gamma_5 F^{\mu\nu}\right)\,\pa_{\mu}\De_\nu
+ \mbox{H.c.}
\eea
These are the free spin-3/2 Lagrangian, and the $\pi N\De$, $\ga N
\De$ couplings respectively. Here $T_a$ are the isospin 1/2 to 3/2 transition matrices
satisfying $T^{\dagger }_a T_b=\frac{2}{3}\de_{ab} -\third
i\veps_{abc}\tau_c$. 

We have kept only the couplings that are linear in the $\De$ field and
lowest order in the pion and the photon fields. In principle, there
are many other couplings ($\pi \pi NN$, $\pi\pi\,N\De$, $\ga\De\De$,
etc.), represented in Eqs.~(\ref{eq:pinlag}) and (\ref{eq:Delag}) by the
dots, that are required by the chiral and electromagnetic-gauge
symmetries. However, they are not relevant for our calculation at the
order we will be considering.

For the purposes of power counting we rearrange the interaction Lagrangian
according to the number of small quantities (momentum, pion mass,
and factors of $e$) that each term carries:
\bea
\lag_I &=& \sum_{i} \lag^{(i)} \, ,\nn\\
\lag^{(1)}&=&   -\frac{g_A}{2f_\pi}\ol N  
(\ga\cdot \pa\, \pi_a) \tau_a \gamma_5 N
+ \frac{i e Q_\pi g_A}{2f_\pi}\ol N  
\ga\cdot A\, \pi \tau \gamma_5 N
+e \,\ol N\,Q_N\,\ga\cdot A\, N +\lag_{\pi N\De} \, ,\nn\\
\lag^{(2)}&=&  \frac{\kappa}{4M_N} 
\ol N\ga^{\mu\nu} N \,F_{\mu\nu} 
+ \half (i e \,Q_\pi \,\pi  A\cdot \pa \pi +\mbox{H.c.}) 
+  e^2 Q_\pi^2 A^2 \pi \pi + \lag_{\ga N\De}^{(g_M)} \, ,\\
\lag^{(3)}&=&  \lag_{\ga N\De}^{(g_E)}\nn \, ,\\
\lag^{(4)}&=&  -\frac{e^2 }{32\pi^2 f_\pi} F_{\mu\nu} \tilde F^{\mu\nu}\pi_3 \, .\nn
\eea

\subsection{Spin-3/2 gauge invariance}

It is important to note that our $N\De$ couplings, 
besides being chiral and gauge invariant, 
are  invariant under the following 
local (gauge) transformation of the spin-3/2 field:
\beq
\eqlab{GI}
\De_\mu(x) \rightarrow \De_\mu(x) +\pa_\mu \eps(x),
\eeq
where $\eps$ is a spinor. This invariance ensures that 
the spin-3/2 field has the correct number of spin degrees of freedom
(i.e., $2s+1=4$), cf.~\cite{Pa98,PaT99}.

As the result of this additional symmetry, 
any vertex involving a Delta field, $\Ga^\mu (p,\ldots)$, 
with $\mu$ being the vector index and $p$ the 4-momentum
of the Delta, will obey the 
{\it transversality condition}:
\beq
\eqlab{trans}
p_\mu\Ga^\mu(p,\ldots)=0.
\eeq
Using the well-known form of the spin-3/2 propagator:
\beq
\eqlab{rsprop}
S^{\mu\nu}(p)=\frac{1}{\slap- M_\De}
\left[-g^{\mu\nu}+\third\ga^\mu\ga^\nu
+\frac{1}{3M_\De}(\ga^\mu p^\nu -\ga^\nu p^\mu)
+ \frac{2}{3M_\De^2} p^\mu p^\nu\right]\,.
\eeq
it is easy to show that the spin-1/2 sector of the RS
propagator decouples~\cite{PaT99} and one may equivalently use the
following propagator:
\beq
\eqlab{s32prop}
\tilde{S}_{\mu\nu} (p)=\frac{-1}{\slap- M_\De}{\mathcal P}^{(3/2)}_{\mu\nu}(p)
\eeq
where 
$   {\mathcal P}^{(3/2)}_{\mu\nu}(p) = g_{\mu\nu} - \frac{1}{3}\gamma_\mu\gamma_\nu
     - \frac{1}{3p^2} (p\hspace{-1.65mm}\slash\gamma_\mu p_\nu
        + p_\mu\gamma_\nu p\hspace{-1.65mm}\slash)
$ is the spin-3/2 projection operator. 

As a matter of fact, it is then also possible to replace
the vertices as follows:
\beq
\tilde{\Ga}_\mu(p,\ldots) =
{\mathcal P}^{(3/2)}_{\mu\nu}  \Ga^\nu(p,\ldots)\,.
\eeq
In this theory $\tilde{\Ga}$ and $\Ga$ are completely equivalent. 
Nevertheless, vertices $\tilde{\Ga}$ are sometimes more
convenient in actual calculations. For example,  the $\pi N\De$
vertex from \Eqref{pinDe}:
\beq
\Ga^\mu(p,k) = (g/M_\De) \ga^{\mu\al\be} {p}_\al k_\be,
\eeq
where $g \equiv h_A/2 f_\pi $ and $k$ is the pion 4-momentum,
can be replaced by
\beq
\tilde{\Ga}_\mu(p,k) = (g/M_\De)\,\slap\, {\mathcal P}^{3/2}_{\mu\nu}(p)\,k^\nu,
\eeq
where we have used:
 $p^\mu {\mathcal P}^{3/2}_{\mu\nu}(p)=0=\ga^\mu {\mathcal P}^{3/2}_{\mu\nu}(p)$ .
Furthermore, using ${\mathcal P}^{3/2}{\mathcal P}^{3/2}={\mathcal P}^{3/2}$
and $ [\slap,{\mathcal P}^{3/2}(p) ]=0$,
 Delta-exchange amplitudes  are computed
effortlessly, e.g.,
\beq
\Ga^\mu(p,k')\,S_{\mu\nu}(p)\,
\Ga^\mu(p,k) = \tilde{\Ga}^\mu(p,k')\,\tilde{S}_{\mu\nu}(p)\,
\tilde{\Ga}^\mu(p,k) = 
\frac{-g^2}{\slap- M_\De}\frac{p^2}{M_\De^2}{\mathcal P}^{(3/2)}_{\mu\nu}(p)\,
 {k'}^\mu k^\nu \,.
\eeq

\subsection{Relation to conventional $\Delta$ couplings}

Our $N\De$ couplings are rather different from the usual ones of,
e.g.~Refs.~\cite{BKM,Ber93,He97,He98,PaS95}.  As a
rule, standard couplings do not have the spin-3/2 gauge
symmetry~\eref{GI}. Exceptions are the $\ga N\De$ coupling of Jones
and Scadron~\cite{JoS73}, which obviously satisfies~\eref{trans},
and the couplings used by Kondratyuk and Scholten~\cite{KoS01}.  We
have adopted the Jones and Scadron convention for the magnetic (M1) coupling,
$g_M$ in \Eqref{ganDe}.

Other conventional couplings, including the $G_1$, $G_2$
representation of the $\ga N\De$ vertex, do not have the spin-3/2
gauge symmetry.  As a result they involve the unphysical lower-spin
sectors of the spin-3/2 field, and hence observables become dependent
on arbitrary ``spin-1/2 backgrounds'' associated with ``off-shell
parameters'' of the Delta. Other pathologies (see Ref.~\cite{Pa98} and
references therein)---all of which can be traced back to the fact that
the couplings violate the degrees-of-freedom-counting constraints of
the free theory---also occur in these theories.

One can establish a relation between the ``inconsistent'' and
``consistent'' couplings using field transformations~\cite{Pa01}, but
this relation holds only in perturbation theory~\footnote{Even then,
it holds only if the ``naive'' Feynman rules apply in the inconsistent
theory, which, strictly speaking, is not true~\cite{Pa98}.} and so is
not strictly applicable when resummation of the Delta contributions is
necessary, as is the case in the computation of the next section.

\section{Compton amplitude in the $\delta$--expansion}

\label{sec-pc}
In Compton scattering the momenta of the particles are characterized
by the photon energy $\w$. For very low photon energies pions can be
``integrated out'' of the theory with all non-analytic effects
associated with their production being replaced by a power series in
$\omega/m_\pi$ (see, for instance Refs.~\cite{pionless}). Clearly, the
condition for this EFT to be effective is $\w \ll m_\pi$. If, instead,
we want to develop an EFT for $\w \sim m_\pi$ we must treat both
$\omega$ and $m_\pi$ as low-energy scales, which means that pions must
appear in the theory as explicit degrees of freedom.

Similarly for the next relevant scale: $\De \equiv M_\De-M_N\approx
293$ MeV. A theory which does not treat it as a low-energy scale is
effective only for $\omega \ll \Delta$. $\chi$PT without explicit
Deltas is an example of such an EFT. There only $m_\pi$ is treated as
a small scale, and it is assumed that $\w\sim m_\pi\ll \De$. To extend
this description to the Delta region, $\w\sim \De$, we need to adopt
$\De$ as a low-energy scale and include the Delta as an explicit
degree of freedom. Thus, we naturally arrive at the scale hierarchy:
\beq
m_\pi\ll \De\ll \La \,.
\label{eq:hier}
\eeq
This hierarchy complies with the assumption of $\chi$PT: $m_\pi\ll
\De$, and so $\chi$PT still gives the dominant effects in the theory
if $\w \sim m _\pi$. At the same time (\ref{eq:hier}) allows us to
extend $\chi$PT to the Delta region.

In developing our power counting below, we will often keep the
dependence of amplitudes on $m_\pi$ and $\Delta$ explicit, so that the
behavior of the amplitudes in the (independent) chiral ($m_\pi
\rightarrow 0$) or the large-$N_c$ ($\De\rightarrow 0$) limits is
manifest~\cite{largeNpapers}.  Nevertheless, for the purposes of
assigning an overall size to the amplitude arising from a particular
graph or set of graphs we would like to have one expansion parameter:
\begin{equation}
\eqlab{coincidence}
\delta \equiv \frac{\Delta}{\Lambda} \approx \frac{m_\pi}{\Delta}\,,
\end{equation}
where we conservatively adopt $\La\approx 600$ MeV, the scale
introduced by the excitation energy of the next baryon resonance. In
fact, $\Lambda$ will represent not only this scale but all of the
various high-energy scales, such as $m_\rho$, $M_N$, $M_\Delta$, and
$4 \pi f_\pi$. Obviously, in this counting, $\De$ scales as $\de$,
while $m_\pi$ scales as $\de^2$.

While $\delta$ is of order one half, the expansion in powers of $\de$,
is, in principle, no worse than $\chi$PT, which is an expansion in
$m_\pi/\De$, or the SSE~\cite{He98}, which is an expansion in powers
of $\Delta/\Lambda$. Note that, since
\Eqref{coincidence} is not necessarily true in worlds with other
values of $N_c$, $m_q$, etc., once that equation is employed, the
connection to the limits $m_q \rightarrow 0$ and $N_c \rightarrow \infty$
is lost unless the chiral and large-$N_c$ limits are taken
simultaneously with  $m_\pi N_c^2$ held fixed.

We assign to each graph an overall $\delta$-counting index, $\al$,
which simply tells us that the graph is of size $e^2
\de^\al/\Lambda$. Because we deal with two different low-energy scales
in our EFT, the index $\al$ has two different expressions, depending
on whether the photon energy $\omega$ is in the vicinity of $m_\pi$ or
$\De$. For a graph with $L$ loops, $N_\pi$ pion propagators, $N_N$
nucleon propagators, $N_\De$ Delta propagators, and $V_i$ vertices of
dimension $i$, the index is
\beq
\eqlab{alphaHB}
\al = \left\{ \begin{array}{cc} 2 \al_{\chi{\mathrm PT}} - N_\De\,, & \w\sim m_\pi; \\
	 \al_{\chi{\mathrm PT}} - N_\De\,, & \w\sim \De, \end{array}\right.
\eeq
where
$ \alpha_{\chi{\mathrm PT}}=\sum_i i V_i -2  + 4 L  - N_N - 2 N_\pi $
is the index of the graph in $\chi$PT without explicit Deltas.

In deriving this power-counting we have used the fact that no graphs
containing vertices with powers of $m_\pi^2$ or $\De$ occur up
to the order to which we work. Such vertices do arise in higher-order
graphs though. In general then, a vertex with $j$ derivatives,
$k$ insertions of the quark mass, and $l$ powers of the $\Delta$-N
mass difference scales as $\w^j\, m_\pi^{2k} \,\De^l$, and so has
overall dimension $i=j+2k+l$. Denoting the number of such vertices by
$V_{jkl}$, the $\de$-index of an arbitrary graph is: $$
\al = \left\{ \begin{array}{ll} \sum\limits_{jkl} (2j+4k+l) V_{jkl} +
2 (4 L  - N_N - 2 N_\pi-2)- N_\De\,, & \w\sim m_\pi;\\
\sum\limits_{jkl} (j+4k+l) V_{jkl} + 4 L  -2 - N_N - 2 N_\pi- N_\De\,, & 
\w\sim \De, \end{array}\right. 
$$
which obviously reduces to Eq.~(\ref{eq:alphaHB}) if only
vertices with $k=l=0$ are present.

\begin{figure}[h,b,t,p]
\vspace*{0.2cm}
\centerline{  \epsfxsize=12 cm
  \epsffile{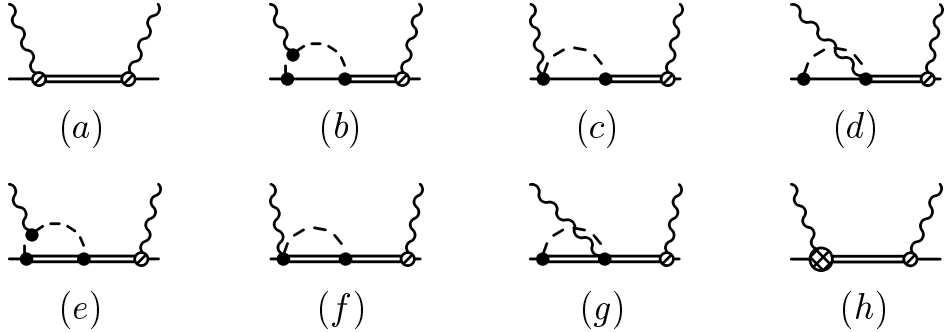}
}
\caption{Eight graphs which are one-Delta-reducible (O$\De$R) and
so become enhanced in the region $\omega \sim \Delta$. The sliced
vertex is the M1 $\ga N \De$ vertex, while the sliced and diced
$\ga N \De$ vertex is the E2 coupling from ${\cal L}^{(3)}$. Solid dots
represent couplings from ${\cal L}^{(1)}$.}
\label{fig-Deltapole}
\end{figure}

In the region $\w\sim\De$, there is an important exception to this
scaling rule: graphs that are one-Delta-reducible (O$\De$R), such
as those in Fig.~\ref{fig-Deltapole}, scale not
as $\de^\al$ but as
\beq
\delta^\alpha \left(\frac{1}{\omega - \Delta}\right)^{N_{O\De R}},
\label{eq:ODRscaling}
\eeq
where $N_\De$ in the equation for $\al$ now counts only the
one-Delta-irreducible propagators, while $N_{O\De R}$ is the number of
\ODR\ propagators. In the low-energy region this does not affect the
power-counting, however in the region $\w\sim \De$ these graphs can be
dramatically enhanced. This forces us to resum all the \ODR\
contributions, which amounts to dressing the Delta propagator, thus
ameliorating the divergence which otherwise occurs at the Delta pole,
and producing a width for the Delta of roughly the
experimentally-observed size.

Details of the dressing are given below. By definition, once dressing
is performed a \ODR\ graph can have only one Delta propagator, and such
a graph then scales as:
\beq
\delta^\alpha \left(\frac{1}{\omega - \Delta -\Si }\right),
\label{eq:ODRscalingnew}
\eeq
where $\Si$ is the self-energy. The expansion for $\Si$ begins at
$\de^3$, and so in the domain $|\w-\De|\sim\de^3$ the 
\ODR\ graphs are enhanced by
$\de^{-2}$ over the value expected from \Eqref{alphaHB}. Thus, the
correct index of a \ODR\ graph in the region $\w\sim\De$ is
\beq
\eqlab{alODR}
\al=\al_{\chi{\rm PT}}-N_\De -2\,.
\eeq 
As a result, for instance, the $s$-channel-pole Delta graph of
Fig.~\ref{fig-Deltapole}(a), which is the simplest \ODR\ graph, is
promoted from NNLO in the low-energy region to LO in the Delta region.

The rest of this section is organized as follows: before giving a
detailed explanation of Compton counting in the $\delta$-expansion we
make a few comments on how our scheme compares to standard HB$\chi$PT
and to the SSE of Ref.~\cite{He97,He98}. We then discuss power
counting for the low-energy region $\omega \sim m_\pi$.  In subsection
\ref{sec-deltadress} we explain the central issue for the
higher-energy domain $\omega \sim \Delta$: the dressing of the Delta
pole. Then in subsection \ref{sec-highpc} we elucidate the impact of
this dressing on the counting for Compton scattering graphs.

\subsection{Comparison with HB$\chi$PT/SSE}

In HB$\chi$PT the Delta is not included as an explicit
``low-energy'' degree of freedom in the Lagrangian. Instead it is
integrated out of the theory, producing a low-energy theory that, in
principle, breaks down for $\omega \sim  \Delta$. Power
counting of graphs is then performed in terms of the index $q$, where
\begin{equation}
q\equiv \frac{\omega}{\Lambda}\sim \frac{m_\pi}{\Lambda}\,,
\end{equation}
where $\Lambda$ is usually assumed to be of order 1 GeV,
although the omission of explicit Deltas  suggests instead
$\Lambda \sim \Delta$.

Hemmert {\it et al.}~\cite{He97,He98} introduced the ``small-scale
expansion'' (SSE) (see also the earlier Ref.~\cite{JM91}), where the
EFT expansion parameter is:
\begin{equation}
\epsilon \equiv \frac{m_\pi}{\Lambda}, \frac{\omega}{\Lambda}, \frac{\Delta}{\Lambda}\,.
\end{equation}
The SSE treats $m_\pi$ and $\Delta$ as the same scale, and hence the
Delta must be included explicitly in both energy domains: $\w\sim
m_\pi$ and $\w\sim \De$. 

This overemphasizes the importance of the Delta somewhat at
low energies.  In contrast, in the low-energy region, the
$\delta$-expansion amplitude is akin to that of HB$\chi$PT.  In the
region $\omega \sim \Delta$ the dressing of the Delta implemented here
is not performed in either HB$\chi$PT---naturally, since Deltas are
``high-energy'' degrees of freedom---or the SSE---as for $\omega \sim
\Delta \sim m_\pi$ all $\pi N$ loop effects are a small correction to the ``bare''
Delta propagator.

The table below summarizes the relationship of the $\delta$-expansion
to HB$\chi$PT and the SSE.

\begin{table}[htb]
\begin{tabular}{|c||c|c|}
\hline
Expansion &  $\quad m_\pi/\La \quad$ & $\quad \Delta/\La \quad$\\
\hline
HB$\chi$PT & $q$        & $1$\\
SSE     & $\epsilon $ & $\epsilon $\\
$\delta$-expansion & $\delta^2 $ & $\delta $\\
\hline
\end{tabular}
\caption{The three different expansion discussed in the text. In all three 
cases the small expansion parameter is of order 1/2, and 
$\Lambda$ is the breakdown scale of the theory.}
\label{table-compare}
\end{table}

\subsection{Power counting for $\omega \sim m_\pi$}

\begin{figure}[h,b,t,p]
\vspace*{0.5cm}
\centerline{  \epsfxsize=12 cm
  \epsffile{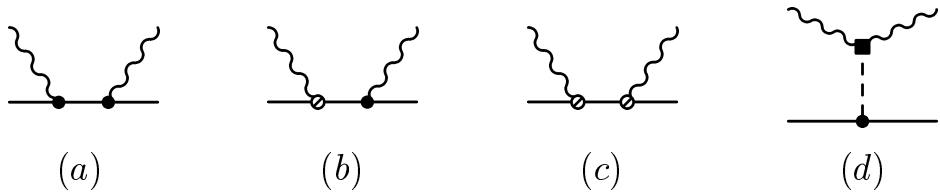}
}
\caption{The four relativistic tree-level graphs without Deltas
that are included in our calculation. (We also include graphs
generated from these graphs by crossing and/or time reversal.)  The
dot is the leading-order $\ga NN$ coupling, while the sliced vertex is
the anomalous-magnetic-moment vertex from ${\cal L}^{(2)}$. The square
indicates the $\pi^0 \rightarrow
\gamma \gamma$ vertex from ${\cal L}^{(4)}$ which 
generates the chiral anomaly.}
\label{fig-graphs}
\end{figure}

Here we make the identification $\omega, m_\pi \sim \delta^2$.
Graphs without Delta propagators then scale exactly as in $\chi$PT,
but with small momenta $q \equiv \delta^2$~\footnote{The electron
charge is usually counted as one power of $q$ in $\chi$PT, and thus
$O(q^3)=O(e^2 q)$ for Compton scattering. Here we do not count the
factor of $e^2$ that is present in all Compton graphs when assessing
the $\delta$-index of a graph.}. The general index of such a graph is
then given by
\Eqref{alphaHB}. The leading contribution then comes from the sum of
the relativistic nucleon Born graph (with $V_1=2$, $N_N=1$,
$L=N_\pi=N_\De=0$) depicted in Fig.~\ref{fig-graphs}(a) and
its crossed partner. Both graphs behave as $e^2/\w\sim
\de^{-2}$ as $\omega \rightarrow 0$. But the divergent parts cancel in the sum,
as the low-energy theorem tells us they must~\cite{LET}. The dominant
term for small $\omega$ is given by the ``Thomson amplitude'':
\begin{equation}
T^{\rm (Th)}=-\frac{({\cal Z} e)^2}{M_N} {\bf \varepsilon}' \cdot {\bf
\varepsilon},
\label{eq:Thomson}
\end{equation}
with ${\bf \varepsilon}$ and ${\bf
\varepsilon}'$ the photon's initial- and final-state polarization
vectors. This, obviously, is $O(\de^0)$.

When the expansion of the relativistic graphs Fig.~\ref{fig-graphs}(a)
and \ref{fig-graphs}(b) in powers of $\omega$ is made there are also
pieces $\sim e^2
\omega=O(\de^2)$. These form part of the NLO
amplitude. The rest of the NLO amplitude is obtained from graphs which
have index $\al=2$: nucleon tree graphs with the anomalous magnetic
moment coupling (i.e., $V_2=2$, $N_N=1$, $L=N_\pi=N_\De=0$), see
Fig.~\ref{fig-graphs}(c), and the $\pi^0$ exchange
graph~\ref{fig-graphs}(d), involving the WZW anomaly, which has
$V_1=V_4=1$, $N_\pi=1$, $L=N_N=N_\De=0$.

\begin{figure}[h,b,t,p]
\vspace*{0.5cm}
\centerline{  \epsfxsize=13.5 cm
  \epsffile{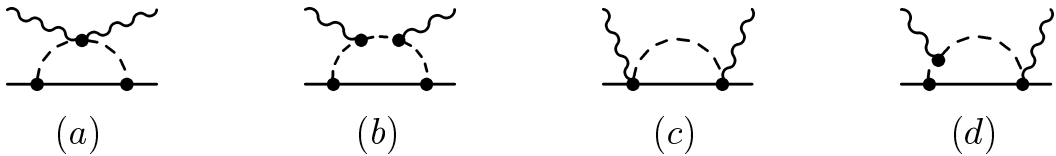}
}
\caption{The dominant $\pi$N-loop contributions to $\gamma$N 
scattering (crossed and time-reversed partners are not shown, but are 
included).}
\label{fig-loopgraph}
\end{figure}

Next we consider $\pi$N-loop contributions to $\gamma$N scattering.
After making a heavy-baryon expansion of our relativistic Lagrangian,
in order to avoid difficulties with the appearance of the scale $M_N$
inside loops~\footnote{The ideal solution to this difficulty would be
to use infrared regularization~\cite{BL99} to compute the $\pi N$
loops. But the result of such a computation should only differ from
the HB$\chi$PT one by terms suppressed by $\delta$. Such terms are
higher order than we are considering here.}, we construct the leading loop
graphs from vertices in ${\cal L}^{(1)}$. This yields the graphs
of Fig.~\ref{fig-loopgraph}, together with their crossed partners, as
reviewed in Ref.~\cite{BKM}. These graphs are specified by $L=1$,
$N_N=1$, $N_\De=0$ and either $V_1=V_2=2$, $N_\pi=3$ or $V_1=2$,
$N_\pi=1$ or $V_1=2$, $V_2=1$, $N_\pi=2$, and hence all have $\al=2$.
They are the only loop graphs with this counting index if we adopt Coulomb
gauge and employ the heavy-baryon expansion. Explicit
computation~\cite{BKM,McG01} reveals that the sum of these graphs
indeed produces a Compton amplitude that behaves as:
\begin{equation}
T^{{\rm (}\pi N~\rm{loop)}}=\frac{e^2}{(4 \pi f_\pi)^2}
\frac{\omega^2} {m_\pi} F^{(1)}\left(\frac{\omega}{m_\pi}\right),
\label{eq:LOloops}
\end{equation}
where $F^{(1)}$ is a non-analytic function whose form is given in
detail for the various possible spin and polarization structures in
Refs.~\cite{BKM,McG01} and in Appendix A. $F^{(1)}$ has the property
that $F^{(1)} \sim 1$ for $\omega \lsim m_\pi$.  A crucial feature of
Eq.~(\ref{eq:LOloops}) is the fact that the sum of these leading loop
graphs is proportional to $\omega^2$. This is also a consequence of
the low-energy theorem~\cite{LET}.

 The counting formula (\ref{eq:alphaHB}) indicates that loop graphs
with insertions from the second-order $\chi$PT Lagrangian ${\cal
L}^{(2)}$ are down by two further powers of $\de$, being of
$O(\delta^4)$. Relativistic corrections to Eq.~(\ref{eq:LOloops}) are
suppressed by $\omega/M$, and so are also $O(\delta^4)$.  Some loop
graphs at $O(\delta^4)$ require renormalization, and the corresponding
counterterms must be included. Meanwhile, graphs with two $\pi$N loops
are $O(\delta^6)$ in this counting. Thus---at least in this energy
domain---it is not until $O(\delta^6)$ that two-pion intermediate
states contribute to the $\gamma$N amplitude. And graphs involving
additional $\pi$N rescatterings are similarly suppressed. Considering
more loops and/or insertions with more derivatives only serves to
further increase the $\delta$-index of graphs. Thus unitarity (in both
the $\pi$N and $\pi\pi$N channels) is violated in our calculation, but
the violation is always an effect of an order beyond that at which we
work.

\begin{figure}[h,b,t,p]
\vspace*{0.5cm}
\centerline{  \epsfxsize=13.5 cm
  \epsffile{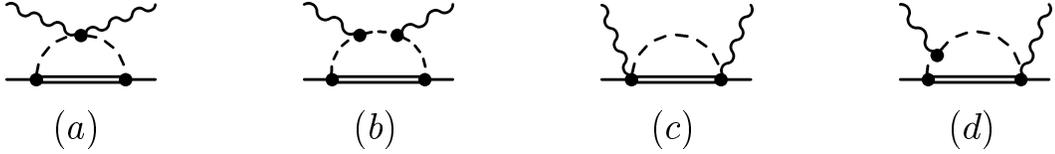}
}
\caption{The dominant $\pi \Delta$-loop contributions to $\gamma$N 
scattering. Again, graphs generated from these by crossing and/or
time-reversal are not shown.}
\label{fig-Deltaloop}
\end{figure}

Graphs containing the Delta begin to contribute at $O(\de^3)$.  The
tree graph with two M1 $\ga N \De$ vertices---see
Fig.~\ref{fig-Deltapole}(a)---has $\al=3$ ($V_2=2$, $N_\De=1$,
$L=N_N=N_\pi$). Meanwhile, the counting of the one-$\pi\De$-loop
graphs, Fig.~\ref{fig-Deltaloop}, is analogous to that for the $\pi N$
loops, the only difference being that now $N_N=0$, $N_\De=1$ instead
of $N_N=1$, $N_\De=0$.  This results in the sum of the graphs in
Fig.~\ref{fig-Deltaloop} being $O(\delta^3)$, i.e. scaling as:
\begin{equation}
T^{{\rm(}\Delta \pi~{\rm loop)}}=\frac{e^2}{(4 \pi f_\pi)^2}
\frac{\omega^2}{\Delta}
H^{(1)}\left(\frac{\omega}{\Delta},\frac{m_\pi}{\Delta}\right),
\label{eq:Deltaloops}
\end{equation}
where $H^{(1)}$ is a non-analytic function which is of order 1 for
$m_\pi/\Delta \sim \delta$, and $\omega/\Delta \lsim 1$. Equation
(\ref{eq:Deltaloops}) is consistent with the low-energy theorem, and
agrees with the explicit computation performed for these loops in
Ref.~\cite{He98}.

In summary: in the $\delta$-expansion the Thomson term is the leading
mechanism for Compton scattering on the nucleon at low energies,
$\omega \sim m_\pi$. In this region pion loops are suppressed by one
power of $\omega \sim m_\pi \sim \delta^2$, exactly as in
HB$\chi$PT. If explicit Deltas are included in the theory the
leading Delta-pole and $\Delta \pi$-loop graphs are suppressed by
$\delta^3$ relative to leading. They thus occur one order higher in
the $\delta$-expansion than the $N\pi$ loop graphs of
Fig.~\ref{fig-loopgraph}. They are, however, still one power of
$\delta^{-1}$ larger than graphs arising from ${\cal L}^{(2)}$
insertions in $\pi N$ loop graphs.

\subsection{Dressing the Delta}
\label{sec-deltadress}

The key issue for the theory in the region $\omega \sim \Delta$ is the
treatment of the Delta pole. One-$\Delta$-reducible (\ODR) graphs
must be resummed in order to remove the divergence
which otherwise occurs when $\slap=M_\Delta$.

Formally, all \ODR\ graphs can be summed via the series:
\beq
S_{\mu\nu}(p) =  S_{\mu\nu}^{(0)}(p) + 
S^{(0)}_{\mu\mu'}(p) \, \Si^{\mu'\nu'}(p) \,S^{(0)}_{\nu'\nu}(p) + \ldots\,, 
\label{eq:resum}
\eeq
where $\Si^{\mu\nu}(p)$ is the full one-Delta-irreducible (\ODI)
Delta self-energy, and $S_{\mu \nu}(p)$ ($S_{\mu \nu}^{(0)}(p)$) is
the dressed (bare) Delta propagator.

\begin{figure}[h,b,t,p]
\vspace*{0.5cm}
\centerline{  \epsfxsize=8 cm
  \epsffile{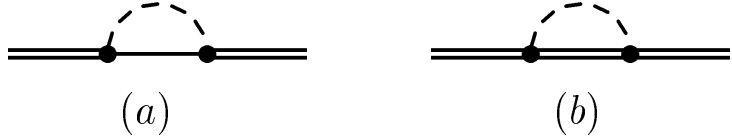}
}
\caption{$\pi$N and $\pi \Delta$ contributions to the $\Delta$
self-energy $\Sigma$. The vertices are from ${\cal L}^{(1)}$
and so both graphs are $O(\delta^3)$.}
\label{fig-selfenergy}
\end{figure}

The function $\Si_{\mu \nu}$ has a $\delta$-expansion of its own:
\begin{equation}
\Si_{\mu \nu}=\Sigma_{\mu \nu}^{(3)} + \Sigma_{\mu \nu}^{(4)} + \ldots.
\end{equation}
This expansion begins at $O(\delta^3)$, with the graphs depicted in
Fig.~\ref{fig-selfenergy}, together with the counterterms
necessary for their renormalization. Insertions from ${\cal L}_{\pi N}^{(2)}$
generate effects in $\Sigma^{(4)}$. These effects include relativistic
corrections to the leading heavy-baryon result
$\Sigma^{(3)}$. Two-loop contributions to the self-energy---including
the leading effect of the $N \pi \pi$ channel---first occur
in $\Sigma^{(5)}$, and are thus smaller by $\delta^2$ than the
dominant piece of $\Sigma$.

If:
\beq
\eqlab{close}
|\omega-\Delta| \sim \delta^3
\eeq
and we keep only the third-order
piece of the self-energy then all terms on the right-hand side of
Eq.~(\ref{eq:resum}) are the same order. A perturbative expansion of
the right-hand side---which is certainly valid for $|\omega - \Delta|
\sim \delta^2$ or larger---is no longer appropriate. Instead, if 
(\ref{eq:close}) holds, the whole series must be resummed, giving:
\beq
S_{\mu\nu}(p)  = \frac{-1}{\slap-M_\De- \bar{\Si}^{(3)}(p)}\, 
{\mathcal P}^{(3/2)}_{\mu\nu}(p)\,.
\label{eq:resummed}
\eeq
\Eqref{close} then defines precisely what we mean by $\omega \sim \Delta$.

In fact the most general Lorentz-covariant form of $\Si^{\mu\nu}(p)$
is rather complicated: it contains up to 10 independent scalar
functions. As a result the dressed propagator does not generally have
the form (\ref{eq:resummed}).  This is a consequence of using
``inconsistent'' spin-3/2 couplings---ones that {\it do not} obey the
symmetry under~\eref{GI}. If, however, couplings which are consistent
in that sense are used then the Delta self-energy can be
written as:
\beq
\Si^{\mu\nu}(p) = \Si(p)\, {\mathcal P}^{(3/2)\mu\nu}(p),
\eeq
with $\Si(p)$ akin to the usual fermion self-energy:
$\Si(p)=A(p^2)\, \slap + B(p^2)$, where $A$ and $B$ are scalar functions.
Dressing then affects only the spin-3/2 piece of the propagator and
the form (\ref{eq:resummed}) results.  The
divergence at $\slap=M_\Delta$ is ameliorated, and no further
resummation is necessary. $\delta$-counting indicates that the effects
of $\Sigma^{(n)}$ for $n \geq 4$ can be included by perturbing
around the propagator (\ref{eq:resummed}).

Delta propagators of this form have been used in other authors'
extensions of chiral perturbation theory to the resonance
region~\cite{unitarizers,Lutz,Tor02}, although in these works it is
not clear why only the spin-3/2 sector is dressed.
Note that in contrast to the work of,
for instance, Ref.~\cite{Lutz}, we do not dress the nucleon pole by
$\pi N$ loops. Arguments analogous to those of this subsection suggest
that nucleon dressing is only necessary from a power-counting point of
view for $\omega \sim 0$, and there $\Sigma(p)$ is purely real. As we
shall now see, after renormalization the real part of $\Si^{(3)}$ 
plays a negligible role in the propagator
(\ref{eq:resummed}).

In Eq.~(\ref{eq:resummed}) the quantity $\bar{\Sigma}^{(3)}$ indicates
that we are resumming the renormalized third-order Delta
self-energy. The explicit renormalization of this quantity will be
performed elsewhere. Here we make a more general argument which
constrains the form and importance of any renormalized self-energy
appearing in Eq.~(\ref{eq:resummed}).

First, observe that the general Lorentz structure of the self-energy
$\Sigma$ results in:
\begin{equation}
S_{\mu \nu}(p)=-\frac{Z(p^2)} {\slap - M(p^2)}\, {\mathcal P}^{(3/2)}_{\mu\nu}
(p),
\label{eq:Sform}
\end{equation}
with $Z$ and $M$ scalar functions of $p^2$. 
After mass, wave function, and coupling constant renormalization these can
be written as:
\begin{eqnarray}
Z(p^2)&=&1 + (p^2 - M_\Delta^2) f_Z(p^2) + i \, \mbox{Im}\,Z(p^2);\\
M(p^2)&=&M_\Delta + (p^2 - M_\Delta^2)^2 f_M(p^2) + i \, \mbox{Im}\,M(p^2),
\end{eqnarray}
with $f_Z$ and $f_M$ real functions of $p^2$. 

Substituting these forms into Eq.~(\ref{eq:Sform}) we find that:
\begin{equation}
S_{\mu \nu}(p)=-\frac{1 + i\,\mbox{Im}\,Z(p^2)}{\slap - M_\Delta - 
i \,\mbox{Im}\,\,M(p^2)} {\mathcal P}^{(3/2)}_{\mu\nu}(p)
+ O\left(\frac{1}{\Lambda}\right).
\end{equation}
In the region (\ref{eq:close}) the pole piece is $O(\delta^{-3})$,
while the remaining ``background'' terms are $O(\delta^0)$. Thus any
corrections to $S_{\mu \nu}(p^2)$ from $\mbox{Re}\,Z(p^2)$ or
$\mbox{Re}\,M(p^2)$ are three powers of $\delta$ beyond leading.
Corrections from $\mbox{Im}\,Z(p^2)$ are equally suppressed, since it too
is $O(\delta^3)$.

Thus, up to corrections which are N$^3$LO, it is sufficient to compute
only $\mbox{Im}~M(p^2)$. Provided that $\omega \leq
\Delta + m_\pi$ this comes exclusively from
Fig.~\ref{fig-selfenergy}(a). A straightforward calculation gives:
\begin{equation}
\mbox{Im}\,M(s) \equiv -\frac{\Gamma(s)}{2}=-\left( 
\frac{h_A}{2 f_\pi}\right)^2 \frac{s+M_N^2-m_\pi^2}
{24\pi M_\De^2} \,{k}^3\,\th(k),
\label{eq:impart}
\end{equation}
where $k$ is the on-shell value of the pion three-momentum: 
\beq
k= \{ [s-(M_N+m_\pi)^2][s-(M_N-m_\pi)^2]/(4s)\}^{1/2} \sim \de.
\label{eq:keq}
\eeq
Thus, the width is $O(\delta^3)$, as promised.

The final form of the resummed
$\Delta$ propagator is then:
\beq
\eqlab{resummed2}
S_{\mu \nu}(p)=-\frac{1}{\slap - M_\Delta + \frac{i}{2}\Gamma(p^2)}
{\mathcal P}^{(3/2)}_{\mu\nu}(p),
\eeq
If this propagator appears in a \ODR\ $\gamma$N graph and
(\ref{eq:close}) is satisfied then it scales as $\delta^{-3}$.

\subsection{Power counting for $\omega \sim \Delta$}

\label{sec-highpc}

The effect of this modified scaling for the Delta propagator is that
(dressed) \ODR\ graphs become the dominant effects for $\omega \sim
\Delta$. Their $\de$-index is given by
\Eqref{alODR}. The Delta-pole graph Fig.~\ref{fig-Deltapole}(a), with
M1 $\gamma N \Delta$ vertices (i.e., $V_2=2$, $N_\De=1$), has $\al=-1$
in the region (\ref{eq:close}) and gives the leading contribution
there. The graph of Fig.~\ref{fig-Deltapole}(h) with one E2-coupling
(i.e., $V_2=1$, $V_3=1$) has $\al=0$, and hence is of next-to-leading
order (NLO) if $\omega \sim \Delta$.

Meanwhile the \ODR\ graphs of Fig.~\ref{fig-Deltapole}(b)--(g) (and
their time-reversed partners) are characterized by $L=N_N=N_\De=1$ and
either $V_1=V_2=N_\pi=2$ or $V_2=N_\pi=1$, $V_1=2$. They too
have $\al=0$ and contribute at NLO. These loop graphs are divergent and
must be renormalized. This is achieved via
graph~\ref{fig-Deltapole}(h). The loop effects may then be included in
the calculation by the use of an energy-dependent E2-coupling: $g_E
\rightarrow g_E(s)$. The leading effect here again arises
from the imaginary part of the loops, hence:
\beq
g_E(s)= g_E +i \left(\frac{g_A h_A}{4 f_\pi^2}\right) 
\frac{s + M_N^2 - m_\pi^2}{24 \pi s} \frac{M_N k^2}{\omega_\gamma}
\frac{2 M_N (M_N + M_\Delta)}{3 M_\Delta \omega_\gamma}
\left[Q_0(\w_k/k) - Q_2(\w_k/k)\right] \th(k) ,
\label{eq:ges}
\eeq
where $\w_\ga= (s-M_N^2)/2\sqrt{s}$, $\w_k=\sqrt{m_\pi^2+k^2}$, while
$k$ is given by \Eqref{keq}, and $Q_l$ is the $l$th Legendre
function of the second kind.

NLO effects can also be obtained by considering corrections from
$\Sigma^{(4)}$ to the leading Delta self-energy $\Sigma^{(3)}$. More
complicated electromagnetic couplings and higher-order terms in
$\Sigma$ lead to effects of NNLO in the $\delta$-expansion of
\ODR\ graphs.

What is the $\delta$-index of graphs without a Delta pole? \ODI\
graphs such as $\Delta \pi$ and $N \pi$ loops obey
Eq.~(\ref{eq:alphaHB})---they are not enhanced. They retain
a positive $\delta$ index, and so are at least NNLO in this
counting.

One might wonder how to reconcile this with Eq.~(\ref{eq:LOloops})
which seems to suggest that the $\delta$-scaling of the dominant $\pi
N$ loops will be, for $\omega \sim \Delta$:
\begin{equation}
e^2 \frac{\omega^2}{m_\pi}=e^2 \frac{\Delta^2}{\Delta^2}=e^2 \delta^0.
\end{equation}
This conclusion is erroneous because $F^{(1)}$ is not $O(1)$ if
$\omega/m_\pi$ is large.  If $\omega/m_\pi \gg 1$ the loop functions
should be expanded about the large-$\omega$ limit, not the small
$\omega$ one, and doing so results in:
\begin{equation}
T^{{\rm (}\pi N~\rm{loop)}}=\frac{e^2}{(4 \pi^2 f_\pi)^2}
\left[c_1 \omega + d_1 m_\pi + c_3 \frac{m_\pi^2}{\omega} + 
c_5 \frac{m_\pi^4}{\omega^3} 
+ \ldots\right].
\end{equation}
Assigning $\omega \sim \Delta$, $m_\pi \sim \Delta^2$ we see that this
series, rather than one in increasing powers of $\omega$ is the
correct one for the ``medium-energy'' regime $\omega \sim \Delta \gg
m_\pi$.  This is completely opposite to a polarizability expansion
in increasing powers of $\omega$. In our approach $\omega
\sim \Delta$ is sufficiently far from threshold that a power-series expansion 
around $\omega=0$ of Delta contributions is not very
enlightening.

\section{In practice}
\label{sec-practice}

\subsection{Defining the NLO calculation}

To perform a complete NLO calculation in the whole energy region,
$0 < \w\lsim \De $, we include all of the nucleon pole and $\pi N$
loop graphs of Figs.~\ref{fig-graphs} and \ref{fig-loopgraph},
together with their crossed partners. To these
we add the \ODR\ graphs of Fig.~\ref{fig-Deltapole}, which  in the low-energy
region contribute at NNLO and above, but for $|\omega - \Delta| \sim \delta^3$ 
give effects of leading and next-to-leading order. We also
include the \ODR\ graph with two $g_E$ $\ga N \De$ couplings, even 
though it is formally NNLO. 

We keep all of these effects in both kinematic regions.  Note that
this means we are always keeping contributions that are, strictly
speaking, beyond the order to which we work. This is done in order to
provide a smooth transition between the two different photon energy
domains. For the same reason, in both regions we always use the
resummed Delta propagator (\ref{eq:resummed2}).

\begin{figure}[h,b,t,p]
\vspace*{0.5cm}
\centerline{  \epsfxsize=11 cm
  \epsffile{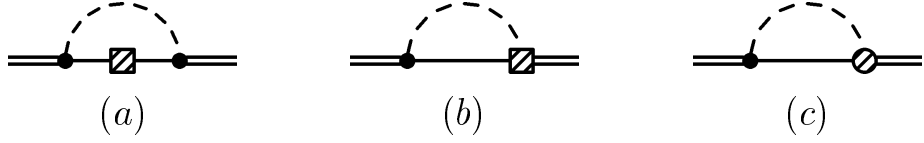}
}
\caption{Heavy-baryon expansion graphs that represent $O(\epsilon^4)$ 
$\pi$N-loop contributions to the $\Delta$ self-energy $\Sigma$. The
sliced squares are vertices from the second-order $\pi N \De$ HB
Lagrangian whose coefficient is fixed by Lorentz invariance. The
sliced circle represents a vertex from ${\cal L}_{\pi N \De}$ with a
coefficient which is {\it a priori} undetermined~(see Ref.~\cite{FM01}).}
\label{fig-selfenergyoq4}
\end{figure}

The power counting of Sec.~\ref{sec-deltadress} indicates that at NLO
we must include effects due to the $O(\delta^4)$ piece of $\Sigma$,
$\Sigma^{(4)}$. The heavy-baryon graphs which contribute to the
$O(\epsilon^4)$ self-energy in a heavy-baryon calculation with
explicit Deltas~\cite{FM01} are shown in
Fig.~\ref{fig-selfenergyoq4}. The relativistic calculation of the
Delta width that led to Eq.~(\ref{eq:impart}) already includes the
effects of the graphs (a) and (b).  As for graph (c), 
this graph gives a contribution to $\Sigma$ which behaves
as~\cite{FM01}:
\begin{equation}
\Sigma^{6(c)}(p) \sim \frac{(b_3 + b_8) h_A}{f_\pi^2}
\int \frac{d^4 k}{(2 \pi)^4} \frac{k^2}{k^2 - m_\pi^2} \frac{v \cdot k}{
v \cdot (p - k)}.
\end{equation}
The imaginary part of this graph is proportional to $\omega \, k \, m_\pi^2$
(with $k$ given by Eq.~(\ref{eq:keq})), so while it is $O(\epsilon^4)$ in
the SSE, in our counting it is $O(\delta^6)$, and so well beyond
the order to which we work. Thus the result (\ref{eq:impart}) is
is already accurate up to corrections of relative order $\delta^2$. 

The $u$-channel Delta-pole graph is NNLO
throughout this entire energy region and is {\it not} included in our
Compton amplitude. Therefore, by looking at its effect on cross
sections we can estimate the importance of NNLO contributions.

\subsection{Covariant decomposition of the Compton amplitude}

To compute the amplitude of Compton-scattering on a spin-1/2
target in a manifestly Lorentz and gauge invariant form, 
we specify it in terms of eight scalar amplitudes
$A_1\ldots A_8$ (instead of the usual six):
\beq
\eqlab{csampl}
M_{fi}= e^2\,\bar u (p')\, 
\sum_{i=1}^{8} A_i(s,t)\,  O_i^{\mu\nu}\, 
{\cal E'}_{\!\!\mu}^{\ast}(q')\, {\cal E}_\nu(q)
\, u(p)
\eeq
where $p'$, $p$ (and $q'$, $q$) are the final and
initial 4-momenta of the nucleon (and photon), respectively;
$u$ is the free nucleon
spinor, ${\cal E}_\mu$ is
a modified photon polarization vector:
\beq
{\cal E}_\mu(q) = \veps_\mu  - \frac{P\cdot \veps}{P\cdot q}\, q_\mu\,,
\eeq
with $P=p+p'$ .
Tensors $O_i$ are given by:
\bea
O_1^{\mu\nu}&=& -g^{\mu\nu} \nn\\
O_2^{\mu\nu}&=& q^\mu {q'}^\nu  \nn\\
O_3^{\mu\nu}&=& - \ga^{\mu\nu}\nn\\
O_4^{\mu\nu}&=& g^{\mu\nu}\,(q'\cdot\ga \cdot q) \nn\\
O_5^{\mu\nu}&=& q^\mu {q'}_{\!\!\al}\ga^{\al\nu} - 
\ga^{\mu\al} q_\al {q'}^\nu \\
O_6^{\mu\nu}&=& q^\mu {q'}_{\!\!\al}\ga^{\al\nu} - 
\ga^{\mu\al} q_\al {q'}^\nu \nn\\
O_7^{\mu\nu}&=& q^\mu {q'}^\nu\,  (q'\cdot\ga \cdot q)\nn\\
O_8^{\mu\nu}&=& i\ga_5 \eps^{\mu\nu\al\be} q_\al' q_\be\nn\,.
\eea
Mandelstam variables $s,\,t,\,u$ are defined as usual:
\bea
s&=& (p+q)^2 =M_N^2 + 2\,p\cdot q\nn\\
t&=& (q'-q)^2 = -2\, q\cdot q' \\
u&=& (p-q')^2 = M_N^2 -2\,p\cdot q'\nn
\eea
where we have used the onshell conditions: $q^2={q'}^2=0$,
$p^2={p'}^2=M_N^2$. Note that $P\cdot q= \half(s-u) = P\cdot q'$.

The representation~\eref{csampl} is obtained by writing down the most
general covariant structure (for the onshell situation) and imposing
the electromagnetic current-conservation condition.  Thus this
representation incorporates both covariance and gauge-invariance in a
manifest way.

The amplitudes $A_i$ are most easily computed in the following
Lorenz-invariant gauge:
\beq
P\cdot \veps =0 = P\cdot \veps'\,.
\eeq
This condition can also be achieved in the Coloumb gauge 
($\veps_0=0=\veps_0'$) by going to the Breit frame: $\vec P=0$. 

In the Coloumb gauge, the structures $O_1$--$O_6$ exactly match the 
ones in the standard decomposition (e.g.~\cite{LET,BKM}), while $O_7$ and $O_8$ 
can be reduced to linear combinations of $O_3$, $O_5$ and $O_6$:
\bea
O_7 &=& (\w^2 +\half t) [-t\, O_3 + O_5] - \w^2 O_6, \\
O_8 &=& \half\w (-t O_3 + O_5 - O_6).
\eea

In particular, the results for the nucleon-exchange
(Born) graphs, the anomaly graph, and the Delta exchange graphs are
specified in terms of the amplitudes $A_1,\ldots,A_8$ in 
Appendix A. As  our calculation of these graphs is fully
relativistic, it differs from that of HB$\chi$PT in Refs.~\cite{BKM,He97}
by terms of $O(\delta)$. 

Meanwhile, for the $\pi N$ loop graphs depicted in
Fig.~\ref{fig-loopgraph}, we have used the $O(q^3)$ HB$\chi$PT result
of Refs.~\cite{BKM,McG01}.  This is done in order to avoid
difficulties with the treatment of the nucleon-mass-scale $M_N$ inside
the loops. The amplitude obtained in a fully-relativistic calculation
of these loops would differ from the result used here by terms of
$O(\omega/M)$ and $O(m_\pi/M)$, i.e. terms down by $\delta^2$. The
loop contributions are given in Appendix A.

We do include one particular relativistic effect because we write the photon
energy $\omega$ that appears in the loop functions as $\omega=\sqrt{s}
- M_N$. The standard choices for $\omega$---c.m. photon
energy~\cite{BKM}, Breit-frame photon energy~\cite{McG01}---differ
from this by terms which are of N$^3$LO in our counting. We adopt this
prescription in order to ensure that the Compton amplitude's $\pi N$
cut occurs at the correct value of $s$.

\section{Results and Discussion}
\label{sec-results}

\subsection{Differential cross sections for $\ga$p scattering}

There are no $\ga$N contact terms at NLO in the $\de$-expansion. This
leaves us with three EFT parameters which must be fixed using the
data: $g_E$, $g_M$, and $h_A$. For $h_A$ we adopt the phenomenological
value $h_A=2.81$ ($f_{\pi N\De}^2/4\pi \simeq 0.35$), which
corresponds to a Delta width $\Ga (M_\De^2) \simeq 111$
MeV---consistent with the range given by the Particle Data
Group~\cite{PDG02}. This value of $h_A$ is roughly 5\% larger than
that obtained from the large-$N_c$ relation
$h_A=\frac{3}{\sqrt{2}}g_A$.

This leaves us with two free parameters: $g_M$ and
$g_E$ of the $\ga N\De$ coupling~\eref{ganDe}.  They represent the
strength of, respectively, the M1 and E2 $\ga N
\rightarrow \De$ transitions.

In principle these are relatively well known from
pion photoproduction. In particular,
their ratio is related to the $R_{EM}=E2/M1$ ratio:
\beq
\eqlab{ratio}
R_{EM} = \frac{ g_E\, \De}{2g_M (M_N+M_\De)
-  g_E\De} \times 100\,\% .
\eeq
The determination of this ratio has recently been the subject of
experimental programs at JLab and MAMI.  The present PDG value is
$R_{EM}=(-2.5\pm 1.0)\%$~\cite{PDG02}. However, this number is measured only
indirectly through the extraction of the ratio of the
pion-photoproduction multipoles at the Delta-resonance position.
These multipoles are affected by a number of background processes and
the relationship (\ref{eq:ratio}) is only strictly true at leading
order in the $\delta$-expansion for the $\gamma$p$\rightarrow\pi$p
amplitude. To fully understand the constraint that pion
photoproduction places on $g_E/g_M$ in this EFT a higher-order
calculation of $\gamma$p$\rightarrow\pi$p using the $\delta$-expansion
is necessary. (For an attempt to compute this process in the SSE see
Ref.~\cite{KC98}.) Absent such a calculation, here we regard $g_M$ and
$g_E$ as free parameters and fit them to get the best agreement with
$\ga$p cross-section data. An important future test of the usefulness
of the $\delta$-expansion will be whether the resultant value for
$g_E/g_M$ is ultimately consistent with that found from pion
photoproduction data using the same framework.

The results for the differential cross section at several different
energies are presented in Figs.~\ref{fig-low} and \ref{fig-high}. The
long-dashed orange curve represents a calculation which includes only
the Born graphs of Fig.~\ref{fig-graphs}.  The dashed blue curve is
the result when the $\pi N$ loops of Fig.~\ref{fig-loopgraph} are
added, and so gives the $O(q^3)$ prediction of HB$\chi$PT.  Finally,
the complete NLO calculation in the $\de$-expansion is represented by
the solid red curve. The sharp rise at backward angles as $\omega$
increases past the pion-production threshold is now reproduced in the
theory. This sharp rise is very difficult to obtain in $\chi$PT
without explicit Deltas.

In Figs.~\ref{fig-low} and \ref{fig-high} we also show
a theoretical error band, demarcated by the red dots. This
is obtained by estimating the size of the NNLO contribution in
the following way:
\beq
T\mbox{(theor.\ err.)} 
= -\frac{e^2}{M_N}\,\veps'\cdot\veps\times \left\{
\begin{array}{ll} \w^2/\De \ , & \w\sim m_\pi \\
\w\, , & \w\sim \De\,. \end{array} \right.
\eeq
The error band in Figs.~\ref{fig-low} and
\ref{fig-high} is plotted with the border between the two kinematic domains
at $\w = 200$ MeV.  The band is an estimate of how far we expect the
NNLO corrections to change the results. It may overestimate the
theoretical error since our calculation already includes a number of
NNLO contributions: relativistic effects coming from tree graphs,
Delta-exchange---which is NNLO for $\omega
\sim m_\pi$---and $\pi N$ loops---which are NNLO for $\w \sim
\Delta$. 

The fit of the cross section favors the following
values for the
$\ga N\rightarrow \De$ parameters:
\beq
\eqlab{gMgE}
 g_M=2.6\pm 0.2 \,,\,\,\,\, g_E=-6.0\pm 0.9 ,
\eeq
The solid line in Figs.~\ref{fig-low} and \ref{fig-high} gives the
result for the central values of $g_M$ and $g_E$. The uncertainty in
Eq.~(\ref{eq:gMgE}) is found by varying these couplings until the
experimental cross section in the Delta region is no longer within
the theoretical error band.    

The resulting value of $g_M$ is
consistent with the large $N_c$ value:  $g_M=
\frac{2\sqrt{2}}{3}(1+\kappa_p)\simeq 2.63$, while the value of $g_E$
is considerably different from phenomenological  expectations. For
instance, from $R_{EM}=-2.5\,\%$ and \Eqref{ratio} one would expect
$g_E=-1$. This problem dramatizes the importance of performing an
analogous calculation of pion photoproduction to check the
consistency of the $\de$-expansion.   

Next we would like to note
that, although formally  the imaginary and real parts of $g_E(s)$ of
Eq.~(\ref{eq:ges}) are the same order, numerically the imaginary part
is smaller by at least a factor of six. $\mbox{Im}\,g_E(s)$ has a
negligible  impact on the angular distributions shown in
Figs.~\ref{fig-low} and
\ref{fig-high}.

The NLO prediction for the differential cross section's
energy-dependence is shown in Fig.~\ref{fig-fixA} for a scattering
angle of $90^o$. The solid purple line is the result when only
$\De$-pole mechanisms are included and $g_M$ and $g_E$ are chosen
according to (\ref{eq:gMgE}). The solid red line gives the result of
our NLO calculation. Individual contributions from Born graphs and
$\pi N$ loops are given by the orange dashed and green short-dashed
lines respectively. The $O(q^3)$ prediction of HB$\chi$PT is
represented by the blue dashed line. This figure shows our NLO
calculation of the Delta width is in good agreement with the data.
This lends support to our adoption of the value $h_A=2.81$.  However,
it must be pointed out that since our EFT was only designed for $0
\leq \omega \lsim \Delta$ the agreement at the higher energies which
is seen in Fig.~\ref{fig-fixA} is probably somewhat fortuitous.

\subsection{Polarizabilities}
The results for the nucleon polarizabilities are worked out in
Appendix B. The leading contribution to polarizabilities is from
$\pi N$ loops and hence is exactly the same as in HB$\chi$PT, see
\Eqref{alphabet}. On the other hand, our results for the Delta contributions
differ from earlier ones in several ways:
\begin{enumerate}
\item First of all, in contrast to the results of Refs.~\cite{Ber93,PaS95},
they are independent of ``off-shell parameters''.

\item 
The leading SSE result of Ref.~\cite{He97} for the magnetic
polarizability is also free of off-shell parameters (in the SSE they
enter at NLO).  Formally, the result of Appendix B agrees with that
found in Ref.~\cite{He97}---apart from a higher-order relativistic
effect which is numerically only of order 10\%: 
$$
\be^{(\De)~{\rm this~work}}_N = \frac{2}{2+\De} \be^{(\De)~{\rm SSE}}_{N}.
$$ 
However, our fit to the cross section prefers a significantly
smaller value of the $\ga N \De$ coupling than was used in
Ref.~\cite{He97} and this leads to a markedly smaller numerical
result for $\be^{(\De)}_p$.

\item Perhaps most importantly, our NLO calculation includes neither 
$u$-channel Delta-exchange nor $\pi\De$ loops, since both are NNLO in
the $\delta$-expansion. In fact, the Delta contributions quoted in
Appendix B stem from both the $s$- and $u$-channel Delta exchange
graphs. However, the NLO calculation presented above has no
$u$-channel graph. Thus to find the polarizabilities that
correspond to the cross-section calculation of the previous
subsection one must halve the Delta pieces of polarizabilities
given in Appendix B.
\end{enumerate}

All of these effects produce a Delta-contribution to the magnetic
polarizability that is significantly smaller than that found in 
previous EFT calculations with explicit Deltas~\cite{But92,Ber93,He97,He98B}:
\beq
\be^{(\De)}_p = \frac{\alpha g_M^2}{(M_N+M_\De)^2 \De} = 
(2.7\pm 0.4) \times 10^{-4}\,\mbox{fm}^3.
\eeq
The Delta contribution to the electric polarizability 
comes solely from the $s$-channel Delta-pole with two $g_E$ couplings:
\beq
\al^{(\De)}_p =- \frac{\alpha g_E^2}{(M_N+M_\De)^3} = 
(-2.0\pm 0.7) \times 10^{-4}\,\mbox{fm}^3.
\eeq

The observant reader will have noticed that the inclusion of $\De$
contributions in the polarizabilities is not, strictly speaking,
consistent at NLO in our power counting. The $s$-channel Delta pole is,
like its $u$-channel counterpart, NNLO for $\w \sim 0$. In spite of
this we have included the Delta contributions in the results shown in
Table~\ref{table-pols}. The numbers there represent the
polarizabilities corresponding to the cross-section calculation
already presented.

We expect that in an NNLO $\de$-expansion calculation the Delta's
effect on $\be_p$ will roughly double. $\al_p$ will also be
modified, thanks to the graphs in Fig.~\ref{fig-Deltaloop} and the
$u$-channel Delta-pole graph with two $g_E$ couplings. Estimating
these effects gives the theory error bars which appear in the second
line of the table. We note that, even though NNLO effects make a
significant difference in the values for $\alpha_p$ and $\beta_p$,
their impact on the low-energy differential cross section is not
large, being represented by the theoretical error band in
Fig.~\ref{fig-low}.  This suggests that the extraction of
polarizabilities from Compton data is a delicate process.


\begin{table}[htb]
\begin{tabular}{||l||c|c||}
\hline\hline
Reference &  \quad $\al_p$ \quad & \quad $\be_p$ \quad \\
\hline
NLO HB$\chi$PT~\cite{BKM} \quad & \quad 12.2 \quad & \quad 1.2~\quad \\ NLO
$\de$ [this work] \quad & \quad $10.2^{+4.2}_{-2.0}$ \quad & \quad
$3.9^{+2.7}_{-0.4}$
\quad\\ 
NLO SSE~\cite{He98B}(\cite{He97}) \quad& \quad 16.4 ($20.8$) \quad&
\quad 9.1 ($14.7$)\quad \\ \hline \hline PDG average~\cite{PDG02}& \quad $12.0\pm 0.7$ \quad
& \quad $1.6\pm 0.6$ \quad \\ LEGS~\cite{LEGS01}& \quad $11.8\pm 2.0$
\quad & \quad $1.4\pm 1.5$ \quad\\ MAMI~\cite{MAMI01} & \quad $11.9
\pm 2.1$ \quad & \quad $1.2\pm 1.4$ \quad \\ Beane {\it et
al.}~\cite{Be02} &
\quad $12.1\pm 1.6$  \quad &
\quad $3.2\pm 1.2$ \quad \\
\hline\hline
\end{tabular}
\caption{Proton polarizabilities in $\chi$PT, the $\de$-expansion, and the SSE,
compared to values extracted from experiment. Results are in units of $10^{-4}$ fm$^3$.}
\label{table-pols}
\end{table}

\section{Concluding remarks}

\label{sec-conc}

The expansion developed in this paper for the $\gamma$N amplitude is
based on the scale hierarchy $m_\pi \ll \Delta \ll \Lambda$. The EFT
expansion parameter employed here is $\delta$, which represents both
the ratio of $m_\pi$ to $\Delta$ and the ratio of $\Delta$ to
$\Lambda$. The success of the resulting ``$\delta$-expansion'' is to
be judged by its efficacy as an EFT of Compton scattering.  

The existence of two different low-energy scales in the theory forces
us to develop independent power countings for the two different
photon-energy regimes. For $\omega \sim m_\pi$ the Compton amplitude
obtained is exactly that of $\chi$PT up to effects from the Delta
which are down by $\delta^3$ relative to leading. On the other hand,
if $\omega \sim \Delta$, \ODR\ graphs dominate the Compton amplitude.
Resummation of the Delta self-energy is necessary in these graphs. The
Delta propagator appearing in them acquires a finite, energy-dependent,
width. Up to effects suppressed by $\delta^3$ this is the only
correction to the Delta propagator that is necessary. Vertex
corrections also appear at NLO in this region.

We performed a calculation of the $\gamma$N amplitude which includes
all effects that are of leading or next-to-leading order in the
region $\omega \sim m_\pi$, as well as all effects that are of
leading or next-to-leading order in the region $\omega \sim
\Delta$. The sum of all of these pieces of the amplitude defines our
NLO calculation. Thus, in each of the two regions considered,
mechanisms of NNLO in that particular region are
included in the calculation. Nevertheless, effects of relative order
$\delta^3$ (e.g. $\Delta \pi$ loops) are omitted in the low-energy
region, while effects of relative order $\delta^2$ (e.g. two-loop
dressing of the Delta) are omitted in the higher-energy domain. These
neglected effects define the theoretical error bar of our calculation.

Note that in this calculation tree-level graphs are computed relativistically,
while for loop graphs we use the HB$\chi$PT result. While this gives
the correct $\gamma$N amplitude up to the order to which we work, a
fully-relativistic treatment of $\pi$N loops would be more
aesthetically pleasing.

After fitting the only two free parameters in our theory, the E2 and
M1 $\ga N \De$ couplings, good agreement with the low-energy $\ga$p
data is found. The spin-independent polarizabilities
$\alpha$ and $\beta$ that result from our NLO calculation are in
reasonable agreement with contemporary extractions from
data~\cite{MAMI01,Be02}. 

The development of the $\delta$-expansion opens up a number of avenues
for further study. Higher-order calculations of Compton scattering on
the nucleon will be necessary to see if the good agreement found at
NLO persists, and if the expansion in powers of $\delta$ is
well-behaved or not. Also, the use of the $\delta$-expansion in other
processes, e.g. pion photoproduction, is an important potential future
application. Indeed, the EFT presented here
should ultimately be judged by its success in simultaneously
describing data on nucleon Compton scattering, pion photoproduction, 
and $\pi$N scattering. Only then can the reliability of the EFT, and hence of 
our extraction of $g_E$ and $g_M$ from $\gamma$N data to NLO in the 
$\delta$-expansion,  really be judged. 

The use of the $\delta$-expansion in two-nucleon systems might also
reap significant rewards. We plan to use the amplitude developed here,
together with consistent two-body currents, in a calculation of
$\gamma$D scattering~\cite{Gr03}. It is also possible that the $\delta$
expansion could be profitably employed to organize the Delta
contributions to the chiral nucleon-nucleon potential developed in
Refs.~\cite{Or96,Ep99,EM01}.

\begin{acknowledgments}
We thank Sergey Kondratyuk and Olaf Scholten for comments on the
manuscript.
D.~R.~P. thanks Harald Grie\ss hammer and Thomas
Hemmert for useful discussions and the Benasque Centre for Science for its
hospitality during part of this work. This research was supported by
the U.~S. Department of Energy under grants DE-FG02-93ER40756,
DE-FG02-02ER41218, and by the National Science Foundation under grant
NSF-SGER-0094668.
\end{acknowledgments}

\appendix

\section{Results for Compton invariant amplitudes}
We define the photon energy as: $
\w = (p\cdot q)/M_N = (s-M_N^2)/2M_N $.
The results for the invariant amplitudes are presented in the
nucleon mass units ($M_N=1$). 
Below $\Z=1$ for the proton, $\Z=0$ for the neutron.
Expressions for the Delta-exchange are obtained using the algebraic
manipulation program Form~\cite{FORM}. 

\subsection{Nucleon $s$-channel:}
\bea
A_1(\w,t) &=& -\frac{1}{2}\left[ \Z^2  
+\mbox{$\frac{1}{4}$}\, (\Z+\kappa)^2 \,t
+ \mbox{$\frac{1}{2}$}\,\kappa^2 ( \w
+\mbox{$\frac{1}{4}$}\, t) \right]\nn\\
A_2(\w,t) &=&-\frac{\kappa}{2\w}\left[
\Z +\mbox{$\frac{1}{2}$}\,\kappa
(1+\mbox{$\frac{1}{4}$}\,\w+\mbox{$\frac{1}{8}$}\,t)\right] \nn \\
A_3(\w,t) &=& A_1(\w,t) \nn\\
A_4(\w,t) &=& -\frac{1}{4\w}\left[ (\Z+\kappa)^2 + 
\mbox{$\frac{1}{2}$}\,\kappa^2\,\w\right]\nn\\
A_5(\w,t) &=& \frac{(\Z+\kappa)^2}{4\w} \\
A_6(\w,t) &=&-\frac{\Z(\Z+\kappa)}{4\w}\nn\\
A_7(\w,t) &=& \frac{\kappa^2}{16\w} \nn\\
A_8(\w,t) &=& - A_4(\w,t) \nn
\eea

\subsection{$\pi^0\rightarrow \ga\ga$ anomaly graph:}
\beq
A_8(\w,t)= \frac{g_A}{(2\pi f_\pi)^2}\frac{2\Z-1}{t-m_{\pi^0}^2},\,\,\,\,
A_1,\ldots, A_7=0.
\eeq

\subsection{Delta $s$-channel:}
\bea
A_1(\w,t) &=& 
F(\w,\De) \left\{ 
\mbox{$\frac{2}{3}$}\, (\w^2+\half\,t) \,(2+\De)\, G_M^2
- \mbox{$\frac{2}{3}$}\,\w^2\,\De\,G_E^2 \right.\nn\\
&+& \mbox{$\frac{2}{3}$}\, \w^3 \,
(G_M^2+G_E^2-G_M G_E) +\mbox{$\frac{1}{8}$}\,t^2\,G_M^2 \nn\\
&+& \left. \w t\, \left[ (2+ \mbox{$\frac{5}{6}$}\,\De 
+ \mbox{$\frac{7}{6}$}\,\w +\quarter\, t) G_M^2
- \mbox{$\frac{1}{6}$}\,(2+\De +\w) G_M G_E
+ \mbox{$\frac{1}{6}$}\,\w\,G_E^2\right]
 \right\}
\nn\\
A_2(\w,t) &=&
F(\w,\De) \left\{ 
- \mbox{$\frac{2}{3}$} \,(2+\De)\, G_M^2
- \mbox{$\frac{1}{3}$} \,\w
\left[ (8+3\De) G_M^2 + 2\De G_E^2 - (2+\De)G_M G_E\right]
\right. \nn\\
&+& \w^2 \,
( - \mbox{$\frac{7}{6}$}\,G_M^2
+ \mbox{$\frac{5}{6}$}\,G_E^2-\mbox{$\frac{2}{3}$}\,G_M G_E) \nn\\
&+& \left. t\, \left[ - \mbox{$\frac{1}{12}$}\,(1-\De)\,G_M^2
- \mbox{$\frac{1}{12}$}\,G_E^2 + \mbox{$\frac{1}{3}$}\,
\w\, (-\half\,G_M^2 + G_E^2 + \mbox{$\frac{5}{4}$}\,G_M G_E)
+\mbox{$\frac{1}{32}$}\,t\,(G_M+G_E)^2
\right]
 \right\}
\nn \\
A_3(\w,t) &=&  F(\w,\De) \left\{ -\third\, \w^2\,
[ 2 G_M^2 + \De\,(G_M^2-G_E^2)] - \third\, \w^3\,
(G_M^2+G_E^2-4 G_M G_E)\right.\nn\\
&+& t \left[ -\mbox{$\frac{1}{6}$}\,(2+\De) G_M^2
+\third\,\w\, G_M\, \left( G_M\,(\mbox{$\frac{3}{2}$}+\De) 
+ G_E\,(2+\De)\right) \right.\nn\\
&+& \left. \left.\w^2\, 
\left( \mbox{$\frac{11}{12}$}\, G_M^2 - \mbox{$\frac{1}{12}$}\, G_E^2
+ \third\, G_M G_E \right) + \quarter\,t\,G_M^2 (\half +\w )  
\right] \right\}\nn\\
A_4(\w,t) &=& F(\w,\De) \left\{-\third\,(2+\De)\, G_M\, (G_M+\w G_E)
- \w\, (1+\third\, \De)\, G_M^2 \right.\nn\\
&+&\left. \third\, \w^2 (G_M^2+G_E^2-G_M G_E)
+ \half\,t\, G_M^2 (\half+\w)\right\}\\
A_5(\w,t) &=& F(\w,\De) \left\{ \third\, (2+\De)\, G_M^2
-\mbox{$\frac{1}{6}$}\,\w \left[
2(1+\De)\, G_M^2 + \De\,G_E^2 + (2+\De)\, G_M G_E\right]
\right.\nn\\
&-& \left. \w^2\,G_M\, (G_M - G_E) 
+ \quarter\, t \left[ G_M G_E +\half\,\De\,G_M\,(G_M+G_E)
-\w\,G_M\,(G_M-G_E)\right] \right\} \nn\\
A_6(\w,t) &=&F(\w,\De) \left\{ \w\,G_M\,[G_M +\half\,\De\,(G_M+G_E)]
+\w^2\,G_M\,(G_M-G_E) \right.\nn\\
&+& \left. \quarter\, t\left[ G_M^2+\half\,\De\,G_M\,(G_M+G_E)
+\w\,G_M\,(G_M-G_E)\right] \right\}\nn\\
A_7(\w,t) &=& F(\w,\De) \left\{ (2+\De)\,
(\mbox{$\frac{7}{12}$}\, G_M^2 +\half\,G_M G_E) 
- \mbox{$\frac{1}{12}$}\,\De\, G_E^2 \right. \nn\\
&+& \left. \w\,\left(\mbox{$\frac{7}{12}$}\, G_M^2 
- \mbox{$\frac{5}{12}$}\,G_E^2 - \third\, G_M G_E \right) 
+ \mbox{$\frac{1}{16}$}\,t\,(G_M+G_E)^2 \right\}\nn\\
A_8(\w,t) &=& F(\w,\De) \left\{ \mbox{$\frac{4}{3}$}\,(2+\De)\,G_M^2
+ \w\,\left[ 4(1+\third\,\De) G_M^2 
-\mbox{$\frac{2}{3}$}\,(2+\De)\,G_M G_E \right]\right.\nn\\
&+& \left. \mbox{$\frac{1}{6}$}\,\w^2 \,(G_M^2+G_E^2-4 G_M G_E)
-\half\, t\,G_M^2\,(\half+\w)\right\} \nn
\eea
with $G_{M,E}=  \frac{3}{2(2+\De)} g_{M,E} $ and
\beq
F(\w,\De)= -\frac{2}{3} \frac{1}{s-M_\De^2+iM_\De \Ga(s)}= -\frac{2}{3}
\frac{1}{2\w-\De(2+\De)+ iM_\De \Ga(\w)}\,,
\eeq
where $\frac{2}{3}=T^\dagger_3 T_3 $ is the isospin factor, and the width
$\Ga$ is given by \Eqref{impart}.

The corresponding $u$-channel graphs are obtained
by crossing so that the crossing-symmetric amplitude is given by
\bea
&& A_i(\w,t)+A_i(-\w',t),\,\,\, \mbox{for}\,\, i=1,2,8 \nn\\
&& A_i(\w,t)-A_i(-\w',t),\,\,\, \mbox{for}\,\, i=3,\ldots ,7 \nn
\eea
with $\w'=p\cdot q'=(1-u)/2$.

\subsection{HB$\chi$PT $\pi N$ loops}

Defining $\barf =(\sqrt{s}-M_N)/m_\pi$ and $\tau =-2{\barf ^2}(1-\cos \theta)$,
where $\theta$ is the center-of-mass angle between the incoming and
outgoing photon momenta, one finds \cite{BKM}:

\begin{eqnarray}
A_1 (s,t)&=& -\frac{g_A^2m_\pi}{8\pi f_\pi^2}
         \left\{ 1- \sqrt{1-\barf^2}
                 +\frac{2-\tau}{\sqrt{-\tau}}
                  \left[\frac{1}{2} \arctan \frac{\sqrt{-\tau}}{2}
                        -I_1(\barf,\tau) \right]\right\},
\nonumber \\
A_2  (s,t)&=& 
        -\frac{g_A^2}{8\pi f_\pi^2 m_\pi}
         \frac{2-\tau}{(-\tau)^{3/2}}
         \left[I_1(\barf,\tau)- I_2(\barf,\tau)\right],
\nonumber \\
A_3  (s,t)&=& \frac{g_A^2m_\pi}{8\pi^2 f_\pi^2}
         \left[ \frac{1}{\barf} \arcsin^2\barf- \barf +2\barf^4
\sin^2 \theta I_3(\barf,\tau)\right],
\nonumber \\
A_4  (s,t)&=& \frac{g_A^2}{4\pi^2 f_\pi^2m_\pi} I_4(\barf,\tau),
\nonumber \\
A_5  (s,t)&=& 
       -\frac{g_A^2}{8\pi^2 f_\pi^2m_\pi}
          [I_5(\barf,\tau)-2\barf^2\cos\theta I_3(\barf,\tau)],
\nonumber \\
A_6  (s,t)&=& \frac{g_A^2}{8\pi^2 f_\pi^2m_\pi}
          [I_5(\barf,\tau)-2\barf^2 I_3(\barf,\tau)],
\label{eq:As}
\end{eqnarray}
where
\begin{eqnarray}
I_1(\barf,\tau) &=& \int_0^1  dz \,
             \arctan \frac{(1-z)\sqrt{-\tau}}{2\sqrt{1-\barf^2 z^2}},
\nonumber \\
I_2(\barf,\tau) &=& \int_0^1  dz \,
             \frac{2(1-z)\sqrt{-\tau (1-\barf^2z^2)}}{4(1-\barf^2 z^2)-\tau (1-z)^2},
\nonumber \\
I_3(\barf,\tau) &=& \int_0^1  dx \, \int_0^1  dz \,
             \frac{x(1-x)z(1-z)^3}{S^3}
             \left[ \arcsin \frac{\barf z}{R}+ \frac{\barf zS}{R^2}\right],
\nonumber \\
I_4(\barf,t) &=& \int_0^1  dx \, \int_0^1  dz \,
             \frac{z(1-z)}{S}\arcsin \frac{\barf z}{R},
\nonumber \\
I_5(\barf,t) &=& \int_0^1  dx \, \int_0^1  dz \,
             \frac{(1-z)^2}{S}\arcsin \frac{\barf z}{R},
\label{eq:Is}
\end{eqnarray}
with
\begin{equation}
S=\sqrt{1-\barf^2 z^2-\tau (1-z)^2x(1-x)}, \qquad R=\sqrt{1-\tau (1-z)^2x(1-x)}.
\end{equation}

\section{Results for polarizabilities}
The nucleon electric ($\al_N$) and magnetic
($\be_N$) polarizabilities:
\bea
\al_N&=& \frac{\al}{2}\frac{\pa^2}{\pa \w^2} A_1^{(NB)}(0,0) 
+ \al\, A_2^{(NB)}(0,0) = 
\frac{5 \pi \al }{6\, m_\pi} \left( \frac{g_A}{4 \pi f_\pi}\right)^2 
- \frac{2 \alpha g_E^2}{(M_N+M_\De)^3}\,,\nn\\
\be_N&=& - \al\, A_2^{(NB)}(0,0) = 2\al \frac{\pa}{\pa t} A_1^{(NB)}(0,0) 
= \frac{\pi \al }{12\, m_\pi} \left( \frac{g_A}{4 \pi f_\pi}\right)^2 
+ \frac{2 \alpha g_M^2}{(M_N+M_\De)^2 \De}\,,
\eea
where $A^{(NB)}_i$ are the amplitudes with
the Born graphs subtracted.

\begin{figure}[h,b,t,p]
\centerline{
  \epsfxsize=13cm
  \epsffile{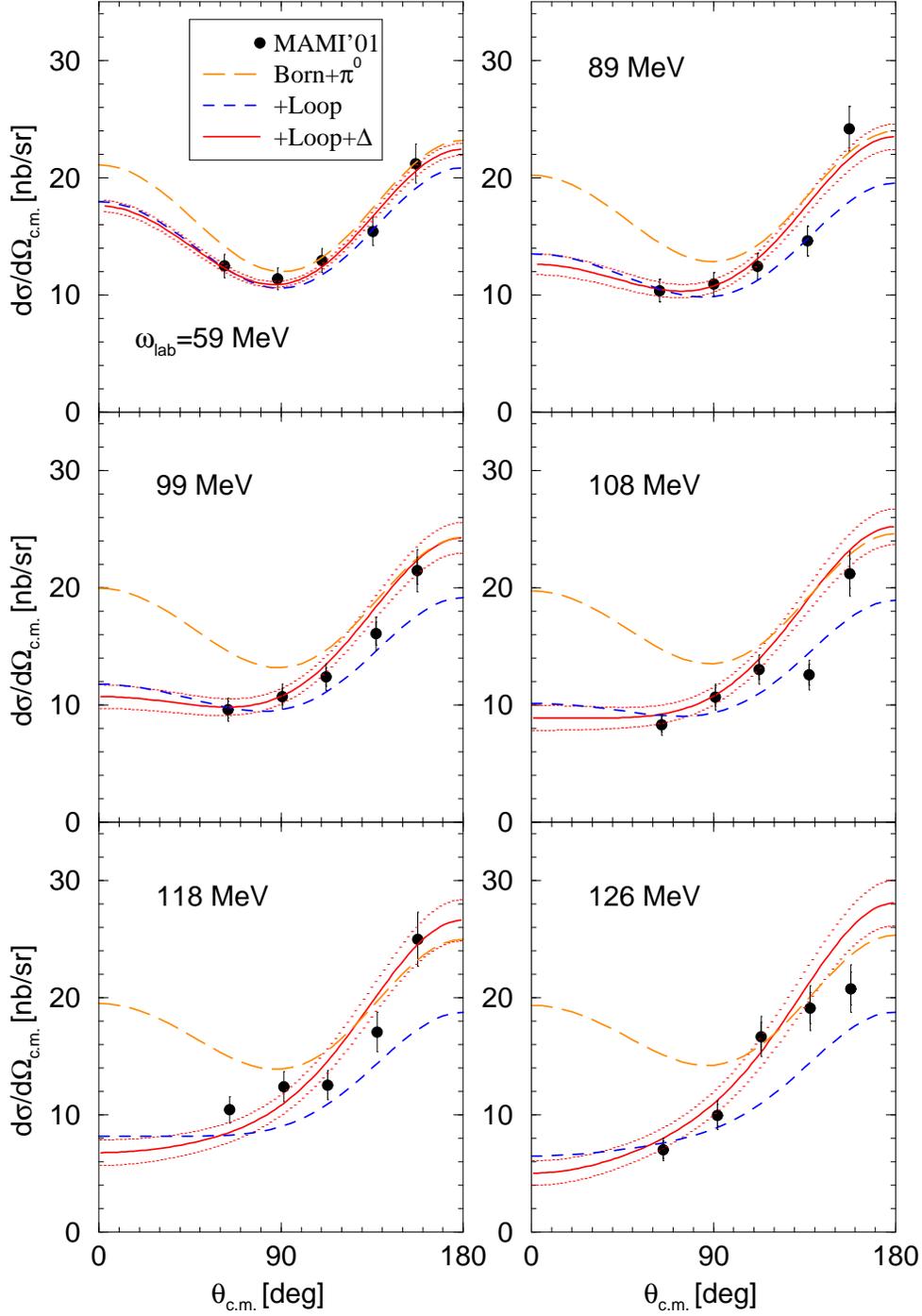}
}
\caption{Angular distribution of the $\ga$p
differential cross section at low energies. Data points are from Ref.~\cite{MAMI01}. The long-dashed orange line represents the sum of nucleon and pion
Born graphs, the blue dashed line gives the NLO $\chi$PT prediction, and
the red solid line is the full result at NLO in the $\delta$-expansion. 
The dots give an estimate of the theoretical error.}
\label{fig-low}
\end{figure} 
\begin{figure}[h,b,t,p]
\centerline{  \epsfxsize=13 cm
  \epsffile{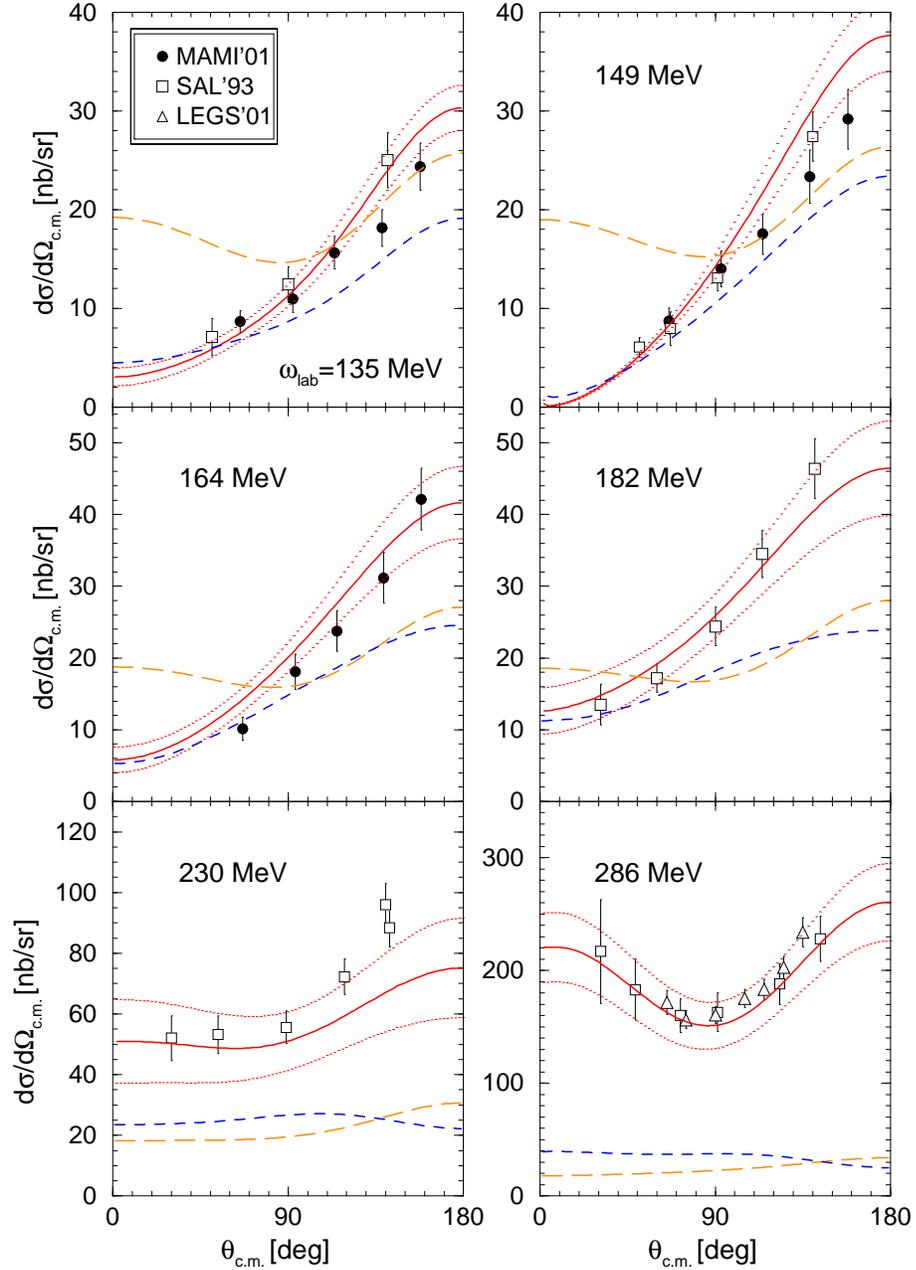}
}
\caption{Angular distribution of the $\ga$p
differential cross section. Legend for the curves is the same as in Fig.~\ref{fig-low}.
Data points are from Refs.~\cite{MAMI01} -- MAMI'01, \cite{Hal93} -- SAL'93,
\cite{LEGS01} -- LEGS'01.}
\label{fig-high}
\end{figure} 

\begin{figure}[h,b,t,p]
\centerline{  \epsfxsize=11 cm
  \epsffile{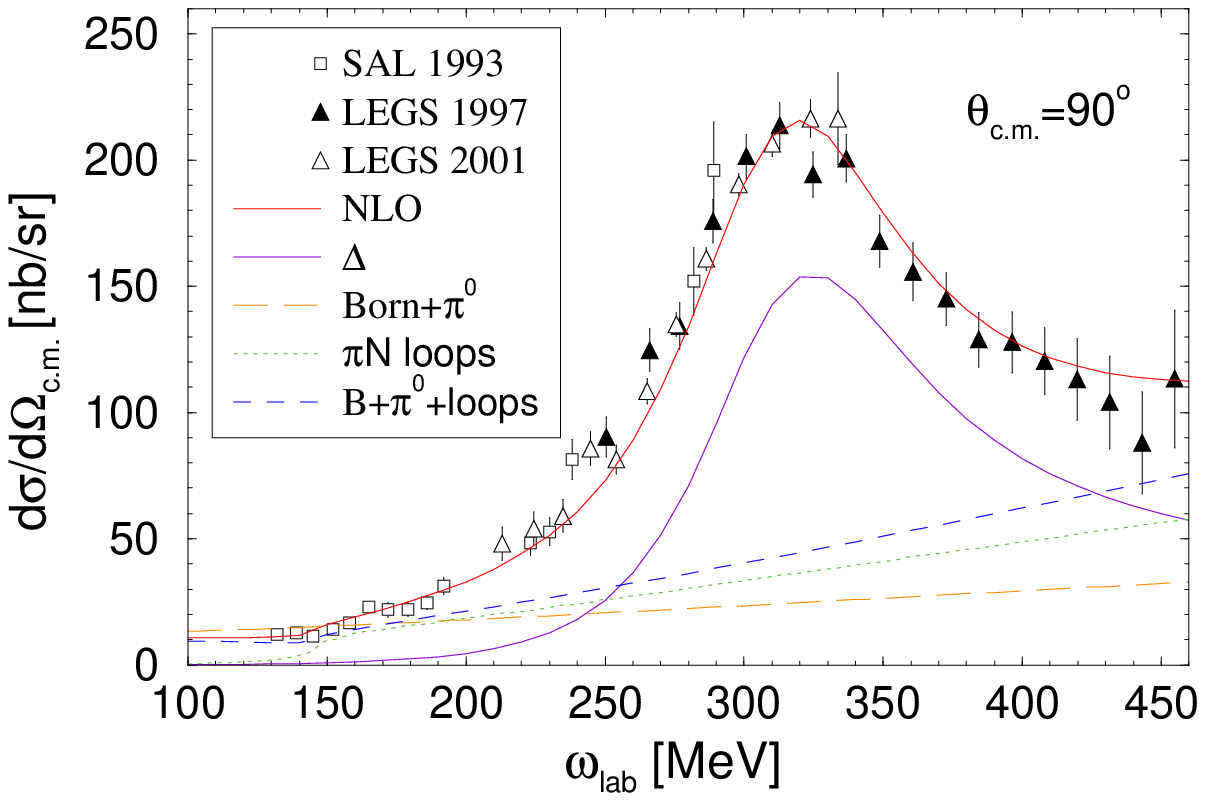}
}
\caption{Energy dependence of the $\ga$p
differential cross section at $90^o$.
Data points are from Refs.~\cite{Hal93} -- SAL'93,
\cite{LEGS97} -- LEGS'97, and \cite{LEGS01} -- LEGS'01.
The red solid, orange long-dashed, and blue dashed lines are as above,
while the purple curve now gives the contribution arising solely from the
Delta, and the green short-dashed curve is the effect of $\pi$N
loops alone.}
\label{fig-fixA}
\end{figure} 


\begin{thebibliography}{99}

\bibitem{Hal93}
E.~L.~Hallin {\it et al.},
``Compton scattering from the proton,''
Phys.\ Rev.\ C {\bf 48}, 1497 (1993).

\bibitem{MacG95}
B.~E.~MacGibbon, G.~Garino, M.~A.~Lucas, A.~M.~Nathan, G.~Feldman and B.~Dolbilkin,
``Measurement of the electric and magnetic polarizabilities of the proton,''
Phys.\ Rev.\ C {\bf 52}, 2097 (1995)
[arXiv:nucl-ex/9507001].

\bibitem{LEGS97}
G.~Blanpied {\it et al.}  [LEGS Collaboration],
``Polarized Compton Scattering From The Proton,''
Phys.\ Rev.\ Lett.\  {\bf 76}, 1023 (1996);
ibid. {\bf 79}, 4337 (1997).

\bibitem{LEGS01}
G.~Blanpied {\it et al.},
``N $\to$ Delta transition and proton polarizabilities from measurements of 
p($\vec{\gamma} , \gamma$), p($\vec{\gamma},\pi^0$), and p($\vec{\gamma},
\pi^+$),'' Phys.\ Rev.\ C {\bf 64}, 025203 (2001).

\bibitem{MAMI01}
V.~Olmos de Leon {\it et al.},
``Low-Energy Compton Scattering And The Polarizabilities Of The Proton,''
Eur.\ Phys.\ J.\ A {\bf 10}, 207 (2001).

\bibitem{Lu94}
M.~Lucas, Ph.~D. Thesis, University of Illinois, unpublished (1994).

\bibitem{Ho99}
D.~L.~Hornidge {\it et al.},
``Elastic Compton scattering from the deuteron and nucleon polarizabilities,''
Phys.\ Rev.\ Lett.\  {\bf 84}, 2334 (2000)
[arXiv:nucl-ex/9909015].

\bibitem{Lu02}
M.~Lundin {\it et al.},
``Compton scattering from the deuteron and neutron polarizabilities,''
arXiv:nucl-ex/0204014.

\bibitem{Ko00}
N.~R.~Kolb {\it et al.}, ``Quasi-free Compton Scattering from the
Deuteron and Nucleon Polarizabilities,'' Phys.\ Rev.\ Lett.\ {\bf 85},
1388 (2000) [arXiv:nucl-ex/0003002].

\bibitem{Ko02}
K.~Kossert {\it et al.},
``Neutron polarizabilities investigated by quasi-free Compton scattering from the deuteron,''
Phys.\ Rev.\ Lett.\  {\bf 88}, 162301 (2002)
[arXiv:nucl-ex/0201015];
``Quasi-free Compton scattering and the polarizabilities of the neutron,''
arXiv:nucl-ex/0210020.

\bibitem{DR}
A.~I.~L'vov, V.~A.~Petrun'kin and M.~Schumacher,
``Dispersion theory of proton Compton scattering in the first and second resonance regions,''
Phys.\ Rev.\ C {\bf 55}, 359 (1997).

\bibitem{DR2}
D.~Drechsel, M.~Gorchtein, B.~Pasquini and M.~Vanderhaeghen,
``Fixed-t subtracted dispersion relations for Compton scattering off the  nucleon,''
Phys.\ Rev.\ C {\bf 61}, 015204 (2000).

\bibitem{PaS95}
V.~Pascalutsa and O.~Scholten,
``On the structure of the $\ga N\Delta$ vertex: 
Compton scattering in the $\Delta$(1232) region and below,''
Nucl.\ Phys.\ A {\bf 591}, 658 (1995);

O.~Scholten, A.~Y.~Korchin, V.~Pascalutsa and D.~Van Neck,
``Pion and photon induced reactions on the nucleon in a unitary model,''
Phys.\ Lett.\ B {\bf 384}, 13 (1996).

\bibitem{Giessen}
T.~Feuster and U.~Mosel,
``Photon and meson induced reactions on the nucleon,''
Phys.\ Rev.\ C {\bf 59}, 460 (1999)
[arXiv:nucl-th/9803057];

G.~Penner and U.~Mosel, ``Vector meson production and nucleon
resonance analysis in a coupled channel approach for energies $m(N) <
\sqrt{s} < 2$ GeV. II: Photon induced results,''
arXiv:nucl-th/0207069.


\bibitem{KoS01}
S.~Kondratyuk and O.~Scholten,
``Compton scattering on the nucleon at intermediate energies and  polarizabilities in a microscopic model,''
Phys.\ Rev.\ C {\bf 64}, 024005 (2001);

``Low-energy Compton scattering on the nucleon and sum rules,''
ibid.  {\bf 65}, 038201 (2002).

\bibitem{Bab97}
D.~Babusci, G.~Giordano and G.~Matone,
``Chiral perturbation theory and nucleon polarizabilities,''
Phys.\ Rev.\ C {\bf 55}, 1645 (1997).

\bibitem{McG01}
J.A. McGovern,
Phys. Rev. C {\bf 63}, 064608 (2001).

\bibitem{Be93} V. Bernard, N. Kaiser, and U.-G. Mei{\ss}ner,
Nucl. Phys. {\bf B383}, 442 (1992);
V. Bernard, N. Kaiser, J. Kambor, and U.-G. Mei{\ss}ner,
Nucl. Phys. {\bf B388}, 315 (1992).

\bibitem{BKM}
V.~Bernard, N.~Kaiser, and U.~G.~Mei{\ss}ner,
``Chiral dynamics in nucleons and nuclei,''
Int.\ J.\ Mod.\ Phys.\ E {\bf 4}, 193 (1995)
[arXiv:hep-ph/9501384].

\bibitem{Be02} S.~R. Beane, M. Malheiro, J.~A. McGovern, 
D.~R. Phillips, and U. van Kolck, ``Nucleon polarizabilities
from low-energy Compton scattering'', 
arXiv:nucl-th/0209002.

\bibitem{But92}
M.~N.~Butler and M.~J.~Savage,
``Electromagnetic polarizability of the nucleon in chiral perturbation theory,''
Phys.\ Lett.\ B {\bf 294}, 369 (1992)
[arXiv:hep-ph/9209204];

M.~N.~Butler, M.~J.~Savage and R.~P.~Springer,
``Strong and electromagnetic decays of the baryon decuplet,''
Nucl.\ Phys.\ B {\bf 399}, 69 (1993)
[arXiv:hep-ph/9211247].

\bibitem{Ber93}
V.~Bernard, N.~Kaiser, U.~G.~Mei{\ss}ner, and A.~Schmidt,
``Aspects of nucleon Compton scattering,'' 
Z.\ Phys.\ A {\bf 348}, 317 (1994) [arXiv:hep-ph/9311354].


\bibitem{He97} T.~R.~Hemmert, B.~R.~Holstein, J.~Kambor,
``$\De$(1232) and the polarizabilities of the nucleon,''
Phys.\ Rev.\ D {\bf 55}, 5598 (1997).

\bibitem{He98}
T.~R.~Hemmert, B.~R.~Holstein and J.~Kambor,
``Systematic 1/M expansion for spin 3/2 particles in baryon chiral  perturbation theory,''
Phys.\ Lett.\ B {\bf 395}, 89 (1997); J.\ Phys.\ G {\bf 24}, 1831 (1998).

\bibitem{He98B}
T.~R.~Hemmert, B.~R.~Holstein, J.~Kambor and G.~Knochlein,
``Compton scattering and the spin structure of the nucleon at low  energies,''
Phys.\ Rev.\ D {\bf 57}, 5746 (1998) [arXiv:nucl-th/9709063].

\bibitem{Gr02} H.~W.~Grie\ss hammer, T.~R.~Hemmert, R.~Hildebrandt, 
and B.~Pasquini, in preparation.

\bibitem{Pa98}
V.~Pascalutsa,
``Quantization of an interacting spin-3/2 field and the Delta isobar,''
Phys.\ Rev.\ D {\bf 58}, 096002 (1998)
[arXiv:hep-ph/9802288].

\bibitem{PaT99}
V.~Pascalutsa and R.~Timmermans,
``Field theory of nucleon to higher-spin baryon transitions,''
Phys.\ Rev.\ C {\bf 60}, 042201 (1999)
[arXiv:nucl-th/9905065].

\bibitem{JM91}
E.~Jenkins and A.~V.~Manohar,
``Baryon Chiral Perturbation Theory Using A Heavy Fermion Lagrangian,''
Phys.\ Lett.\ B {\bf 255}, 558 (1991).


\bibitem{JoS73}
H.~F.~Jones and M.~D.~Scadron,
Ann.\ Phys.\ {\bf 81}, 1 (1973).

\bibitem{Pa01}
V.~Pascalutsa,
``Correspondence of consistent and inconsistent spin-3/2 couplings via  the equivalence theorem,''
Phys.\ Lett.\ B {\bf 503}, 85 (2001)
[arXiv:hep-ph/0008026].

\bibitem{pionless}
J.~W.~Chen, H.~W.~Griesshammer, M.~J.~Savage and R.~P.~Springer,
``Gamma deuteron Compton scattering in effective field theory,''
Nucl.\ Phys.\ A {\bf 644}, 245 (1998);

H.~W.~Griesshammer and G.~Rupak,
``Nucleon Polarisabilities from Compton Scattering on the Deuteron,''
Phys.\ Lett.\ B {\bf 529}, 57 (2002).

\bibitem{largeNpapers}
R.~Flores-Mendieta, C.~P.~Hofmann, E.~Jenkins and A.~V.~Manohar,
``On the structure of large N(c) cancellations in baryon chiral  perturbation theory,''
Phys.\ Rev.\ D {\bf 62}, 034001 (2000);

T.~D.~Cohen,
``Chiral and large-$N_c$ limits of quantum chromodynamics and 
models of the baryon,''
Rev.\ Mod.\ Phys.\ {\bf 68}, 599 (1996) [arXiv:hep-ph/9512275];

T.~D.~Cohen and W.~Broniowski,
``The Role of the delta isobar in chiral perturbation theory and hedgehog soliton models,''
Phys.\ Lett.\ B {\bf 292}, 5 (1992);

W.~Broniowski and T.~D.~Cohen,
``Response of nucleons to external probes in hedgehog models. 1. Electromagnetic polarizabilities,''
Phys.\ Rev.\ D {\bf 47}
, 299 (1993);

T.~D.~Cohen,
``Electromagnetic properties of the Delta in the large N(c) and chiral  limits,''
arXiv:hep-ph/0210278.

\bibitem{LET}
F.~E. Low,  Phys.\ Rev.\ {\bf 96}, 1428 (1954);
M.~Gell-Mann and M.~L.~Goldberger,
{\it ibid.} 1433 (1954).

\bibitem{BL99}
T.~Becher and H.~Leutwyler,
``Baryon chiral perturbation theory in manifestly Lorentz invariant form,''
Eur.\ Phys.\ J.\ C {\bf 9}, 643 (1999)
[arXiv:hep-ph/9901384].

\bibitem{unitarizers}
E.~Oset, E.~Marco, J.~C.~Nacher, J.~A.~Oller, J.~R.~Pelaez, A.~Ramos
and H.~Toki,
``Photoproduction of meson and baryon resonances in a chiral unitary  approach,''
Prog.\ Part.\ Nucl.\ Phys.\ {\bf 44}, 213 (2000);

U.~G.~Meissner and J.~A.~Oller,
``Chiral unitary meson baryon dynamics in the presence of resonances:  Elastic pion nucleon scattering,''
Nucl.\ Phys.\ A {\bf 673}, 311 (2000);

A.~Gomez Nicola, J.~Nieves, J.~R.~Pelaez and E.~Ruiz Arriola,
``Improved unitarized heavy baryon chiral perturbation theory for $\pi$N  scattering,''
Phys.\ Lett.\ B {\bf 486}, 77 (2000).

\bibitem{Lutz}
M.~F.~Lutz and E.~E.~Kolomeitsev,
``Relativistic chiral SU(3) symmetry, large N(c) sum rules and meson  baryon scattering,''
Nucl.\ Phys.\ A {\bf 700}, 193 (2002)
[arXiv:nucl-th/0105042].

\bibitem{Tor02}
K.~Torikoshi and P.~J.~Ellis,
``Low energy pion nucleon scattering in the heavy baryon and infrared  schemes,''
arXiv:nucl-th/0208049.

\bibitem{PDG02}
Particle Data Group, ``Review of Particle Physics,'' Phys.\ Rev.\ D
{\bf 66}, 010001 (2002).

\bibitem{FM01}
N.~Fettes and U.~G.~Meissner,
``Pion nucleon scattering in an effective chiral field theory with  explicit spin-3/2 fields,''
Nucl.\ Phys.\ A {\bf 679}, 629 (2001)
[arXiv:hep-ph/0006299].

\bibitem{KC98}
C.~W.~Kao and T.~D.~Cohen,
``The pion photoproduction in the Delta(1232) region,''
Phys.\ Rev.\ C {\bf 60}, 064619 (1999)
[arXiv:nucl-th/9811062].




\bibitem{Gr03} H.~W.~Grie\ss hammer, T.~Hemmert, V.~Pascalutsa and D.~R.~Phillips,
in progress.

\bibitem{Or96} C.~Ordonez, L.~Ray and U.~van Kolck,
``The Two-Nucleon Potential from Chiral Lagrangians,''
Phys.\ Rev.\ C {\bf 53}, 2086 (1996)
[arXiv:hep-ph/9511380].

\bibitem{Ep99} E.~Epelbaum, W.~Glockle and U.~G.~Meissner,
``Nuclear forces from chiral Lagrangians using the method of unitary  transformation. II: The two-nucleon system,''
Nucl.\ Phys.\ A {\bf 671}, 295 (2000)
[arXiv:nucl-th/9910064].

\bibitem{EM01} 
D.~R.~Entem and R.~Machleidt,
``Accurate nucleon nucleon potential based upon chiral perturbation  theory,''
Phys.\ Lett.\ B {\bf 524}, 93 (2002)
[arXiv:nucl-th/0108057].

\bibitem{FORM}
J.~A.~M. Vermaseren, ``Symbolic Manipulation with FORM, version 2,''
Computer Algebra Nederland, Amsterdam, 1991.

\end{thebibliography}
\end{document}